x
\documentclass[natbib]{svjour3}                     
\pdfoutput=1

\smartqed  
\usepackage{tabularx}
\usepackage{graphicx}
\usepackage{hyperref}
\usepackage{color}
\usepackage{mathtools}
\usepackage{amsfonts}
\usepackage{amssymb}
\usepackage{booktabs}
\usepackage{tabu}
\usepackage{ulem}
\usepackage{textcomp}
 \journalname{Space Science Reviews}
\newcommand{\aap}{{Astron. Astrophys.}}
\newcommand{\apj}{{Astrophys. J.}}
\newcommand{\apjl}{{Astrophys. J. Lett.}}
\newcommand{\apjs}{{Astrophys. J. Suppl.}}
\newcommand{\aj}{Astron. J.}
\newcommand{\epsl}{{Earth Planet. Sci. Lett.}}

\newcommand{\mnras}{MNRAS}

\newcommand{\pss}{{Planet. Space Sci.}}
\newcommand{\araa}{Ann. Rev. Astron. Astrophys.}
\newcommand{\ssr}{Space Sci. Reviews}
\begin{document}

\title{The composition of the protosolar disk and the formation conditions for comets
}


\author{K. Willacy         \and
        C. Alexander,        \and
        M. Ali-Dib,            \and
        C. Ceccarelli,        \and
        S. B. Charnley,          \and
\\
        M. Doronin,          \and
        Y. Ellinger,            \and
        P. Gast,                \and
\\
        E. Gibb,                 \and
        S. N. Milam,              \and
        O. Mousis, 	           \and
\\
        F. Pauzat,             \and
        C. Tornow,             \and
        E. S. Wirstr\"om       \and
        E. Zicler
}


\institute{K. Willacy \at
              MS 169-506, NASA Jet Propulsion Laboratory, California
              Institute of Technology, 4800 Oak Grove Drive, Pasadena, CA
              91109, USA,
              Tel.: +44-818-354-3467,
              Fax: +44-818-354-8895
              \email{karen.willacy@jpl.nasa.gov}           
           \and
           C. Alexander \at
           MS 321-590, NASA Jet Propulsion Laboratory, California
           Institute of Technology, 4800 Oak Grove
           Drive, Pasadena, CA 91109, USA
           \and
           M. Ali-Dib \at
           Institut UTINAM, CNRS/INSU, UMR 6213, Universit{\'e} de
           Franche-Comt{\'e}, 41 bis Avenue de l'Observatoire, BP1615,
           25010 Besan\c{c}on, France
           \and
           C. Ceccarelli \at
           $^{1}$Univ. Grenoble Alpes, IPAG, F-38000 Grenoble, France 
           \at
           $^{1}$CNRS, IPAG, F-38000 Grenoble, France
           \and
           S. B. Charnley \at
           Astrochemistry Laboratory, Mail Code 691, NASA Goddard
           Space Flight Center, Greenbelt, MD 20771, USA
           \and
           M. Doronin$^{1,2}$ \at
           $^1$Sorbonne Universit{\'e}, UPMC Univ Paris 06, CNRS UMR, 7616, Laboratoire de Chimie Th{\'e}orique (LCT), 75252 Paris CEDEX 05, France
           \at
           $^2$Sorbonne Universit{\'e}, UPMC Univ Paris 06, CNRS UMR, 7092, Laboratoire de Physique Mol{\'e}culaire pour l'Atmosph{\`e}re et l'Astrophysique (LERMA/LPMAA), 75252 Paris CEDEX 05, France
           \and           
           Y. Ellinger \at
           Sorbonne Universit{\'e}, UPMC Univ Paris 06, CNRS UMR, 7616, Laboratoire de Chimie Th{\'e}orique (LCT), 75252 Paris CEDEX 05, France
           \and
           P. Gast \at
           Institute of Planetary Research (DLR), Rutherfordstra{\ss}e
           2, 12489 Berlin, Germany
           \and
           E. Gibb \at
           Department of Physics and Astronomy, University of
           Missouri -- St. Louis, St. Louis, MO 63121
           \and
           S. N. Milam \at
           Astrochemistry Laboratory, Mail Code 691, NASA Goddard
           Space Flight Center, Greenbelt, MD 20771, USA
           \and
           O. Mousis \at
           Aix-Marseille Universit\'e, CNRS, LAM (Laboratoire
           d'Astrophysique de Marseille) UMR 736, 13388, Marseille, France.
           \and
           F. Pauzat \at
           Sorbonne Universit{\'es}, UPMC Univ Paris 06, CNRS UMR, 7616, Laboratoire de Chimie Th{\'e}orique (LCT), 75252 Paris CEDEX 05, France
           \and    
           C. Tornow \at
           Institute of Planetary Research (DLR), Rutherfordstra{\ss}e
           2, 12489 Berlin, Germany
           \and
           E. S. Wirstr\"om \at
           Department of Earth and Space Sciences, Chalmers University
           of Technology, Onsala Space Observatory, SE-439 92, Onsala, Sweden
           \and
           E. Zicler \at
           Sorbonne Universit{\'es}, UPMC Univ Paris 06, CNRS UMR, 7616, Laboratoire de Chimie Th{\'e}orique (LCT), 75252 Paris CEDEX 05, France
}

\date{Received: date / Accepted: date}

\maketitle

\begin{abstract}

  Conditions in the protosolar nebula have left their mark in the
  composition of cometary volatiles, thought to be some of the most
  pristine material in the solar system.  Cometary compositions
  represent the end point of processing that began in the parent
  molecular cloud core and continued through the collapse of that core to
  form the protosun and the solar nebula, and finally during the
  evolution of the solar nebula itself as the cometary bodies were
  accreting.  Disentangling the effects of the various epochs on the
  final composition of a comet is complicated.  But comets are not the
  only source of information about the solar nebula.  Protostellar
  disks around young stars similar to the protosun provide a way of
  investigating the evolution of disks similar to the solar nebula
  while they are in the process of evolving to form their own solar
  systems.  In this way we can learn about the physical and chemical
  conditions under which comets formed, and about the types of
  dynamical processing that shaped the solar system we see today.

This paper summarizes some recent contributions to our understanding
of both cometary volatiles and the composition, structure and
evolution of protostellar disks. 

\end{abstract}

\section*{Abbreviations}
\begin{tabular}{ll}
AGB & asymptotic giant branch \\
ALMA & Atacama Large Millimeter Array\\
CAI & calcium-rich aluminium-rich inclusion\\
FUN & Fractionated and unknown nuclear isotopic effects\\
ISM & Interstellar medium\\
ISRF & interstellar radiation field \\
IRAM & Institut de Radioastronomie Millim\'etrique telescope\\
MRI & magneto-rotational instability\\
MHD & magnetohydrodynamical \\
OPR & ortho-to-para ratio\\
PSN & protosolar nebula\\
SLR & short-lived radionuclide\\
Gaseous inner disk & Inside the iceline \\
Inner disk & Inside 35 AU\\
Outer disk & Outside 35 AU
\end{tabular}

\section{Introduction}


Cometary volatiles are some of the most pristine of solar system
materials, having remained relatively unprocessed since the comets
formed in the inner solar system some 4.2 billion years ago.
Consequently they retain in their composition signatures of the
chemical and physical conditions under which they formed.  
Observations of cometary comae
provide the mixing ratios or relative abundances of various
species and have demonstrated that these can vary greatly between
comets, e.g., \cite{ahearn12} and \cite{mumma} and references therein, 
with little correlation with dynamical family.  The
origin of these compositional variations is not well understood but
must lie in the physical and chemical processes that were active at
the time and location at which the cometary volatiles were formed.  To
understand these processes requires not only a study of the comets
themselves but also of protostellar disks similar to the one from which
our solar system formed.  Observations of protostellar disks allow 
for the important disk processes to be investigated while they are
occurring and can test models of the evolution of the early solar system.

Some key questions that can potentially be answered by
interdisciplinary investigations involving both protostellar disk
modeling and observations, and the observations of comets are:
\begin{enumerate}
\item Can synergy between protoplanetary disk modeling and
  compositional studies of comets inform us about the chemical history
  and processes in the early solar disk? 

\item What is the origin of the diversity of observed cometary compositions?

\item Where did comets form in the solar nebula?

\item Are D/H ratios in volatile molecules and spin temperatures
  cosmogonic indicators for comets? If so, what can they tell us about
  the formation conditions in the early solar system? 

\item What is the range of D/H ratios in cometary water and can we
  infer the degree to which comets contributed volatiles to the early
  terrestrial planets? 

\item Can cometary volatile abundances be used to constrain the extent
  of radial and vertical mixing in the early protoplanetary disk? 
\end{enumerate} 

This paper summarizes some recent developments in our understanding of
cometary compositions and their relationship to the conditions under
which the comets formed.
The paper is arranged as follows.  Section~\ref{sec:karen} covers some 
observational and modeling studies of protostellar disks.  
Section~\ref{sec:fractionation} discusses fractionation processes in
protostellar disks including using deuterium fractionation as a tracer
of the links between comets, interstellar chemistry and nebular chemistry.
Section~\ref{sec:n} considers the
origin of the nitrogen deficiency in comets and Section~\ref{sec:spin}
discusses spin temperature as a cosmogenic indicator in comets.
Section~\ref{sec:na} presents a possible explanation for the
observations of neutral sodium cometary tails.  Finally, Section~\ref{sec:summary}
summarizes this paper and briefly discusses the potential impact of
Rosetta on our understanding of the formation conditions of comets.

\section{\label{sec:karen}Overview of the structure and composition of protostellar disks}
Protostellar disks form during the
star formation process.  A molecular cloud core collapses under gravity
forming a protostar in the center.  As material continues to fall
inwards conservation of angular momentum results in the formation of a
disk around the central protostar.  In the early stages, the disks are
gas-dominated with masses $\sim$
1\% of the mass of the protostar. They rotate about the protostar with a Keplerian velocity
profile.   Disks contain a wide range of physical conditions, ranging
from cold ($\sim$ 10 -- 20 K in the outer disk midplane) to hot (with
gas temperatures up to several thousand K in the surface layers).
Irradiation conditions also vary, with the surface layers experiencing
the full force of the stellar UV field, but with little radiation
reaching the midplane.  Consequently there are a wide variety of
chemical conditions depending on location within the disk.  The star
and disk formation process, and the chemistry and dynamics of disks
are discussed in more detail in recent reviews by
\cite{dutrey14,pont14} and \cite{turner14}.

Disks can be very turbulent places.  Turbulence is difficult to detect
directly but evidence for its effects can be seen in the presence of 
small dust grains high above the disk midplane.  Without turbulence these
grains would quickly be removed from these high altitudes by settling
\citep{dd04}.   The dynamical motions also drive collisions between
grains, which can lead to both grain growth, eventually forming
planetesimals, and to creation of new small grains or dust.  With
grain growth, the grains will eventually reach a size at which their
motions decouple from those of the gas and  they begin to settle under
the influence of gravity to the 
midplane, where they can undergo further collisions and
growth to form planetesimals.  The removal of grains from the surface
layers allows the stellar UV to penetrate deeper into the disk,
changing its composition.  Turbulence can also drive chemical changes
by bringing together material processed in different parts of the disk
and conversely chemistry can affect the dynamics by controlling the
ionization levels.  Consequently dynamics and chemistry strongly
influence each other (see Section~\ref{sec:turb}).

In recent years protostellar disks have been an area of active
research, with their molecular content, dust properties, and
physical conditions being studied in detail. The
observational work has been coupled with simulations that can
interpret the observations and make predictions about those regions of
the disks which are not amenable to direct observational study.  This
synergy of the observations and modeling work has  revealed a wealth
 of information about the chemistry and physics of 
disks, and in turn has informed
our understanding of the early history of our own solar system.
This section reviews some of the recent observational and modeling
work related to protostellar disks, concentrating on the early stages
of evolution when the disk is still gas-dominated (the T Tauri phase).

\subsection{The structure of a protostellar disk}
\begin{figure}[!tbh]
\centering
\includegraphics[width=0.32\linewidth]{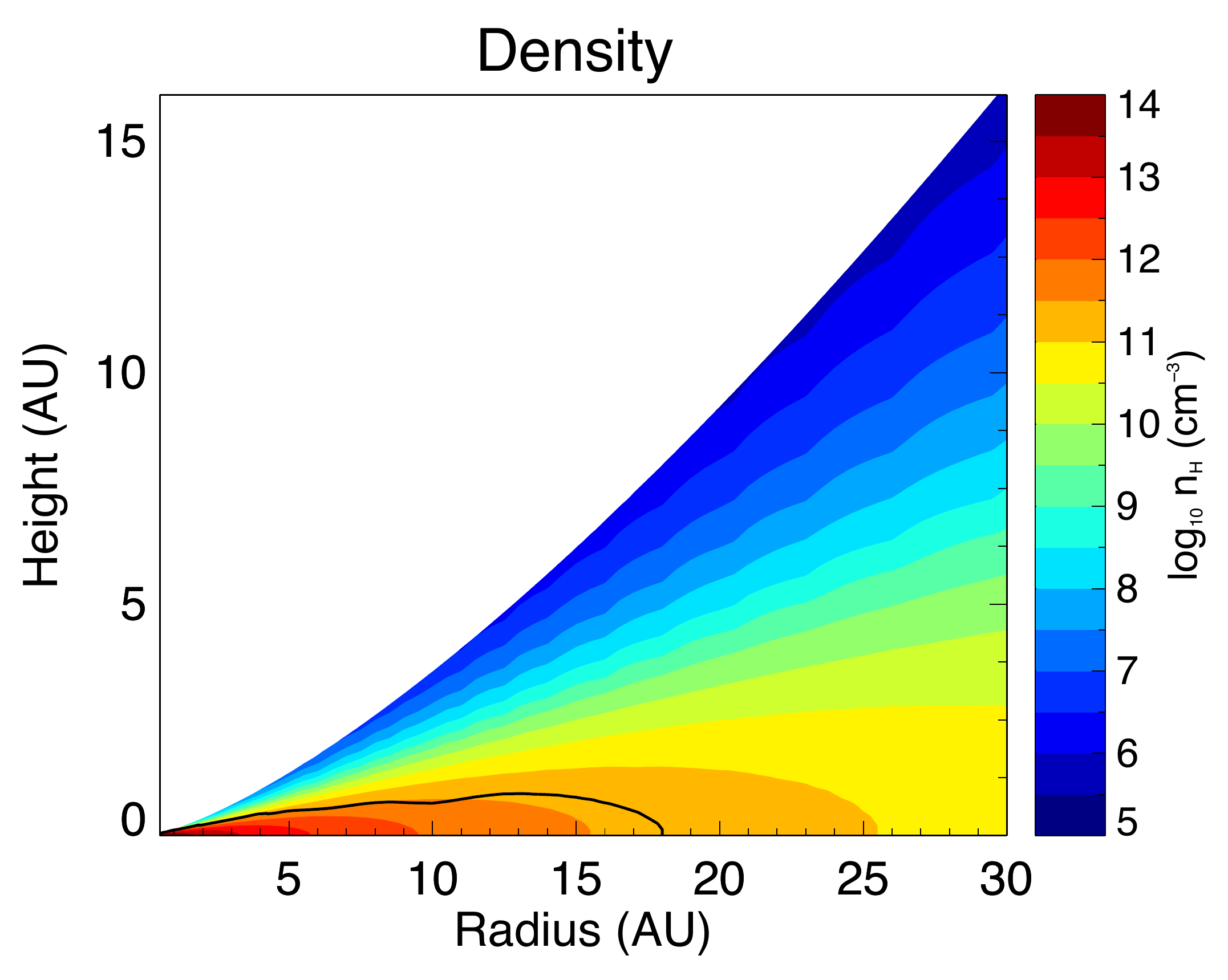}
\includegraphics[width=0.32\linewidth]{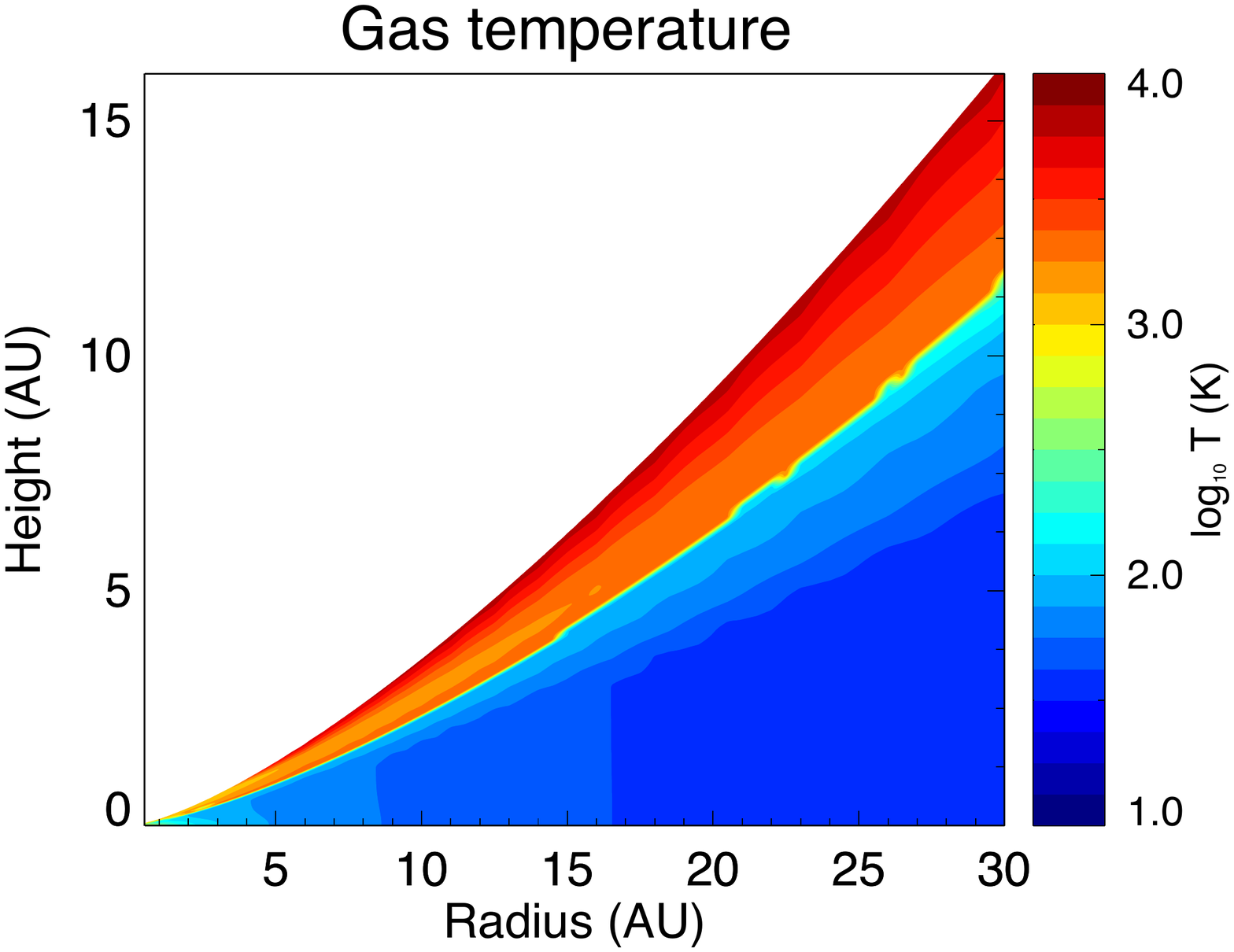}
\includegraphics[width=0.32\linewidth]{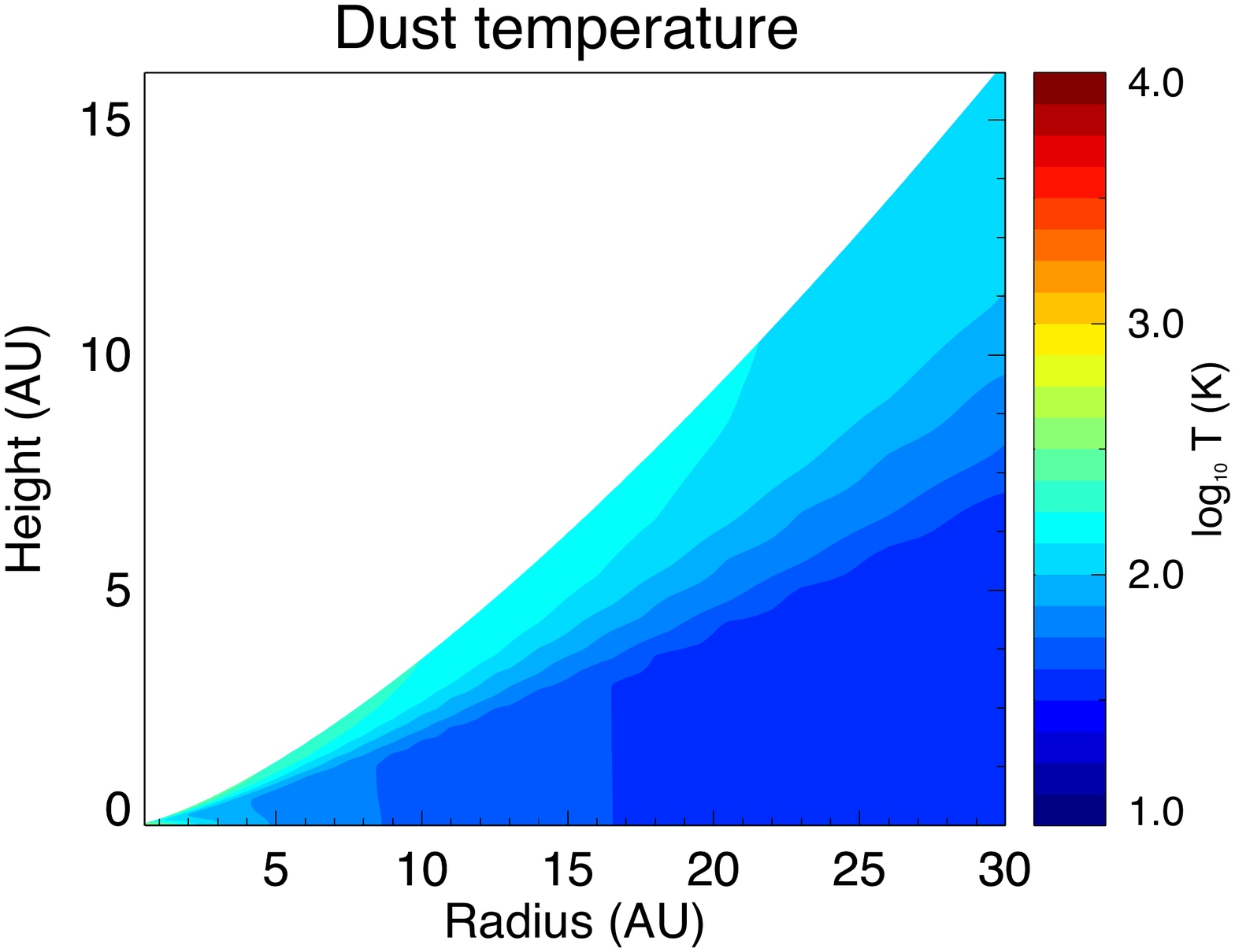}
\caption{\label{fig:disk}The physical structure of a typical T Tauri
  disk.  
  Shown are the density ({\it left}), gas temperature ({\it center})
  and dust temperature ({\it right}). The density and dust temperatures
  are calculated by a hydrodynamical model of Paola D'Alessio with a
  similar mass and surface density profile to the minimum mass solar nebula.
The gas temperature is determined via thermal balance and is coupled
to the dust temperature throughout most of the disk, but is much
  hotter in the surface layers due to the photoelectric effect.  The
  black contour line on the density plot
  indicates the likely location of the dead zone, determined by the
  location where the ionization level x(e) $<$ 10$^{-12}$.  In this
  region the turbulent motions driven by MRI will be severely
  inhibited.  }
\end{figure}

Protostellar disks are characterized by strong radial and vertical
gradients in temperature and density (Figure~\ref{fig:disk}).
The stellar UV field controls the temperature and hence the chemistry
throughout much of the disk.  The penetration of the UV into the disk is governed by the
opacity which is mainly provided by dust absorption.  In the
surface layers, where the UV field is unattenuated, 
the photoelectric effect can efficiently heat the gas to much higher
temperatures than the dust.  Close to the midplane, where the UV
field is reduced, the dust and grain temperatures are coupled. 

\begin{figure}[!tbh]
\includegraphics[width=\linewidth]{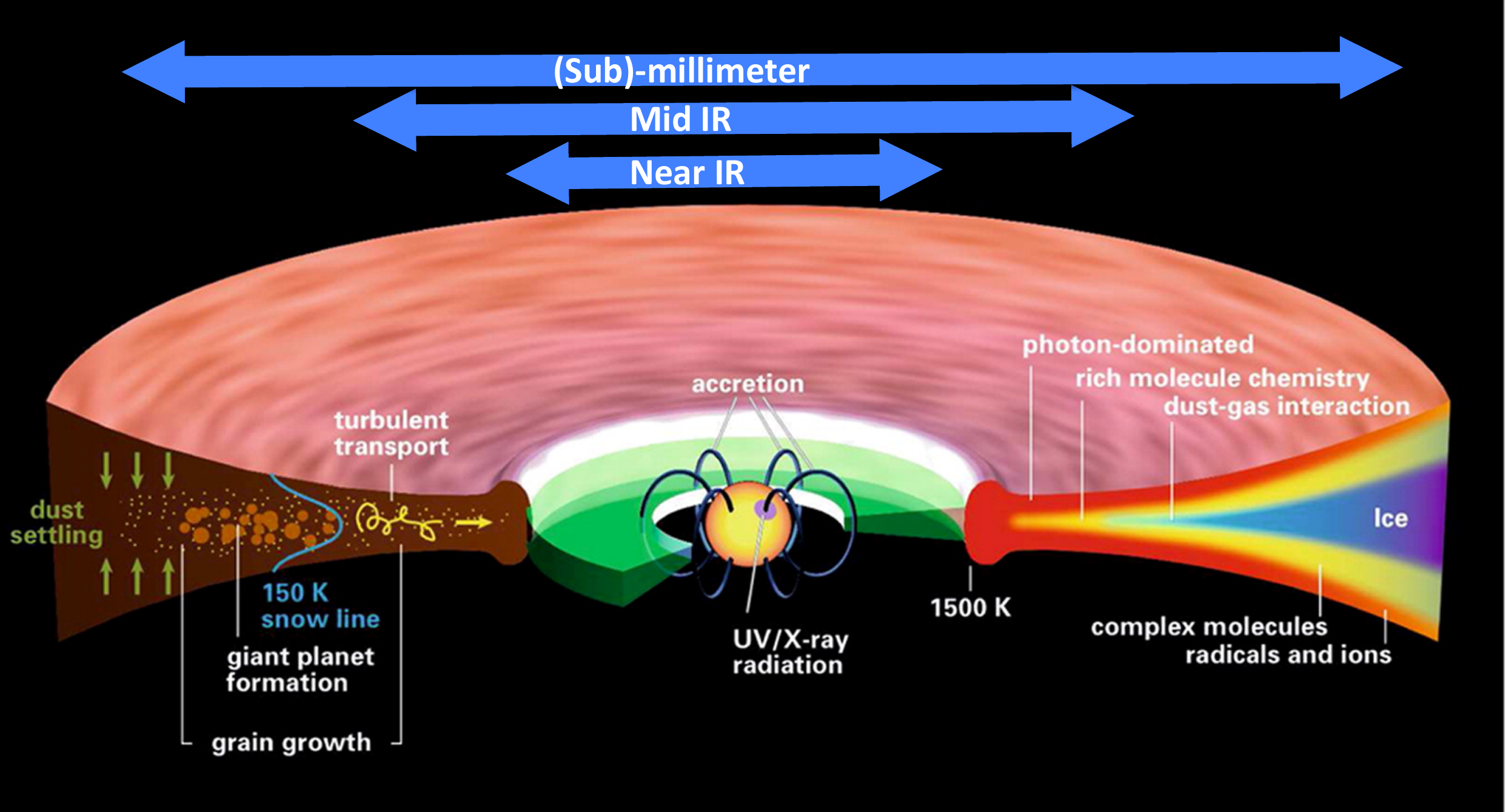}
\caption{\label{fig:disk_structure}The physical and chemical
  environment of a 1 -- 5 Myr old protostellar disk around a Sun-like
  star.  Reprinted from
  \cite{hs13} with permission. \textcopyright (2013) American Chemical Society.}
\end{figure}

These gradients in physical conditions lead to a wide variety of
chemical environments (Figure~\ref{fig:disk_structure}).  Models have
found that outside of the water ice line the disk can be divided vertically
into three chemically distinct layers. At the surface, the disk is
dominated by the effects of UV.  Ices are desorbed (either thermally
or by photodesorption) and efficient photodissociation means that the
gas is mainly composed of atoms and their ions.  In the midplane most
molecules are frozen out onto the surfaces of dust grains.   
Between these two layers is a region where the grains are warm enough
to allow at least some molecules to thermally desorb and which is
shielded enough that photodissociation is inefficient, allowing the
molecules to survive once in the gas.  It is this warm molecular layer
that is detected in most gas phase observations of the outer regions
of protostellar disks (see Section~\ref{sec:diskobs}).

As the distance to the protostar decreases the disk warms up.  In the
midplane, the ice mantles will begin to desorb, with different
molecules being released at different radii, depending on their
binding energies -- the most volatile species being released at larger
radii than the less volatile ones.  This results in a series of ice
lines, which mark the transition from ice to vapor for
each molecule (see Section~\ref{sec:snowlines}).  The icelines are
curved, and in the outer disk are almost horizontal with the molecule
existing as ice below this line and as gas above it
\citep[see][]{pont14}.  Once inside the water iceline
(within a few AU of the protostar) the molecules are all in the gas
phase.

Figure~\ref{fig:disk_structure} also illustrates some of the main
physical processes that are active in disks.  In particular,
turbulence is a crucial process, enabling dust coagulation and driving
chemical changes by mixing material processed under different physical
conditions within the disk.  What drives this turbulence is still a matter of
debate but one of the main candidate processes is the
magneto-rotational instability \citep[MRI; e.g.,][]{bh91,hb91,hawley96}
which can efficiently transport angular momentum in ionized accretion
disks.  A minimum level of ionization is required \citep[in the
minimum mass solar nebula the ionization fraction, x(e), must be $>$
10$^{-12}$ at 1 AU;][]{is05}.  This can be difficult to achieve in the
shielded midplane regions of a protostellar disk, where the high
column densities prevent UV and cosmic rays from penetrating.  Hence
there are regions in the disk where the MRI is not active.  These are
known as ``dead zones'' and they are predicted to occur throughout
much of the midplane in the planet-forming region.  The approximate
location of the dead zone in the disk shown in Figure~\ref{fig:disk}
is indicated by the black contour line in the density plot.

The dependence of MRI-driven turbulence on the ionization means that
there is a close link between the chemistry and the dynamics.  The
chemistry determines the ionization level and therefore the strength
of the turbulence, but the dynamical motions themselves can drive
chemical changes through mixing.  This in turn affects the ionization
and consequently the dynamics
\citep[e.g.,][]{is05,in08,in06,turner07}.  Hence the evolution of
chemistry and dynamics are inextricable linked.

\subsection{\label{sec:diskobs}Observations of protostellar disks}
Protostellar disks have been observed at wavelengths from the UV to
the millimeter.  Each wavelength can probe a different region of the
disk.  The majority of information about the molecular composition of
disks comes from the infrared and the (sub)-millimeter.  The disk is
mostly optically thick in the infrared, therefore these observations
are restricted to the warm gaseous region inside of the iceline or to the surface layers.  At these
wavelengths gaseous molecules can be observed in both absorption and
emission.  Ices can also be observed in absorption if illuminated by
the central star in an edge-on disk, e.g., \cite{pontoppidan}, but the
spectra can be difficult to interpret due to contributions from
material outside of the disk.  Near- and mid-infrared observations are
most sensitive to the inner regions of disks, while the far-infrared
can probe the surface layers at larger radii (see Henning and Semenov
(2013).

Spitzer, operating in the mid-infrared, has shown that inside of the
iceline the disk
is chemically rich, with the terrestrial planet-forming region
containing organics such as C$_2$H$_2$, HCN, as well as
CO, CO$_2$, water and OH \citep[e.g.,][]{cn08,salyk08,pont10}.  The detected
molecules are the result of a rich gas phase molecular chemistry, in a
region where temperatures range from a few hundred kelvin up to 1000
K, and densities are high ($>$ 10$^{8}$ cm$^{-3}$).

Spitzer observations of the gaseous inner disk have also been used to make
inferences about the conditions beyond the iceline.  For example,
\cite{najita13} found the HCN/H$_2$O ratio inside of the iceline to be
correlated with disk mass, suggesting that this ratio may reflect
changes to the C/O ratio induced by the formation of icy planetesimals
outside of the iceline.  Outside of the iceline much of the oxygen
is tied up in water ice, whereas the carbon is mainly in the gas phase
in the form of more volatile molecules such as CO.  Planetesimals
outside of the iceline will form from grains coated with water ice.  Since the
planetesimals do not move about the disk as easily the small
grains and the gas, this will trap the oxygen outside of the iceline
and the material that moves inwards to the inside of the iceline will be
enriched in carbon.  Consequently the gaseous inner disk will be enriched in
organics compared to water.  Since more massive disks are likely to
form planetesimals more efficiently than their lower mass counterparts
they will have a higher C/O ratio inside of the iceline.

In the far-infrared, Herschel has detected a number of molecules
including cold water in the outer disks of TW Hya \citep{hogerheijde}
and DG Tau \citep{podio}.  In both cases the water emission comes from
a region of the disk where grain temperatures are low enough for this
molecule to be expected to be frozen onto dust grains.  Its presence
in the gas can be explained by photodesorption driven by the stellar
UV field \citep{wl00,dominik}. \cite{hogerheijde} found that the
predicted flux from photodesorbed water to be higher than observed,
and suggested that the answer to this discrepancy could be grain
coagulation. As the grains grow they decouple from the gas motions and
sink towards the midplane, taking their ices with them.  This removes
water ice from the molecular layer, leaving less to be photodesorbed,
and thus reducing the gas phase abundance.

Emission lines from some molecules in the disk can be optically thin in the (sub)-millimeter and so in theory
these wavelengths can probe the whole of the disk.  In practice,
however, resolution constraints restrict these observations to $\gtrsim$ 30
-- 100 AU from the protostar, although ALMA allows observations to
within a few AU of the closest protostars e.g.\ \citep{ALMA15}.  Observations by both
interferometers and single dish facilities have detected the emission
from molecules in the warm molecular region and their distributions
can be mapped using interferometers. .  Relatively
simple molecules have been observed including CO (and its
isotopologues), H$_2$CO, CS, C$_2$H, HCN, CN, HNC, DCN, HC$_3$N,
c-C$_3$H$_2$, N$_2$H$^+$, HCO$^+$, DCO$^+$, H$_2$D$^+$
\citep[for a recent review see][]{dutrey14}.
The fractional abundances relative to H$_2$ are
generally lower than observed in molecular clouds because of the
combined effect of depletion in the cold dense midplane and
photodissociation in the surface layers.  Both processes reduce the gas
phase molecular abundances.

In addition to the composition of the
disk, the detection of molecules can also provide information about
the physical conditions under which they exist and the chemical
processes that are active (Table~\ref{tab:molecules}).  Observations
of molecular deuteration are especially important since they trace the
thermal history of the disk (see also Section~\ref{sec:deut}).
Deuterated molecules, with their lower column densities, also provide
a tracer of the gas in the midplane, e.g., ALMA observations of DCO$^+$
have been used to trace the location of the CO iceline in a Herbig
AeBe disk \citep{mathews13} and H$_2$D$^+$ is a potential tracer of
the ionization level in the midplane \citep{ceccarelli04}.


\begin{table}[!htb]
\begin{center}
\bgroup
\def\arraystretch{1.5}
\begin{tabular}{ll}
\hline
Process & Molecules \\
\hline
Temperature & CO, $^{13}$CO, C$^{17}$O, C$^{18}$O, H$_2$CO, H$_2$O, OH, H$_2$\\
Density & CO, H$_2$CO, HCO$^+$, HC$_3$N\\
Ionization & HCO$^+$, N$_2$H$^+$, CH$^+$\\
Velocity & CO, CS, HC$_3$N\\
Photodissociation & CN, HCN, HNC, OH, H$_2$O, C$_2$H\\
Photodesorption & H$_2$O, OH\\
Grain surface chemistry & H$_2$CO, complex organics, CO$_2$, C$_2$H$_2$ \\ 
High temperature chemistry & Complex organics\\
Deuteration & HD, DCO$^+$, DCN, H$_2$D$^+$\\
\hline
\end{tabular}
\egroup
\caption{\label{tab:molecules}Molecules observed in protostellar disks
  and the processes they trace.}
\end{center}
\end{table}

\subsection{Protostellar disk modeling}

Chemical models of protostellar disks vary in their complexity, but
they all predict the three-layer structure beyond the iceline as shown in
Figure~\ref{fig:disk_structure}.  The chemical simulations require as
their basis, a background disk model to provide the density,
temperature and UV field as a function of location in the disk.  These
can be provided by hydrodynamical models of a 1+1D steady-state
$\alpha$-disk model in vertical hydrostatic equilibrium such as those
of
\cite{dalessio05}\footnote{\url{http://www.cfa.harvard.edu/youngstars/dalessio/}};
(see also Figure~\ref{fig:disk}).  These physical disk models are computationally
intensive  and are are generally run independently
of the chemical simulations.  Current chemical models are based on
detailed chemical kinetics and can contain hundreds of species (in
both the gas and the ices) and thousands of reactions.  These models
include the ionizing effects of UV photons, cosmic rays, X-rays and
short-lived radionuclides such as $^{26}$Al.  Some models have also
focused on isotopic chemistry such as deuterium
\citep[e.g.,][]{ah99,ww07,willacy_woods09,kavelaars11,furuya13,albertsson14}
or carbon and oxygen isotopes \citep[][see also
Section~\ref{sec:c_frac}]{woods_willacy09}.  The simplest models
assume that the conditions in the disk did not change with time and
ignore dynamical motions, although a few also consider advection e.g.,
\cite{willacy98,aikawa99,ww07,woods_willacy09,willacy_woods09}.  More
recent models have included processes such as grain growth and
settling, e.g., \cite{an06,fogel11,vasyunin11,akimkin13},
dynamical motions,
\cite[e.g.][]{ilgner04,willacy06,semenov06,sw11,in06,turner06,aikawa07,hersant09,heinzeller11}
and hydrodynamical evolution of the disk structure itself 
\citep[e.g.][]{dr,yu14}.  Here we highlight a few recent results.

\subsubsection{Formation of complex organics in protostellar disks}

\begin{figure}[htb]
\includegraphics[width=\linewidth]{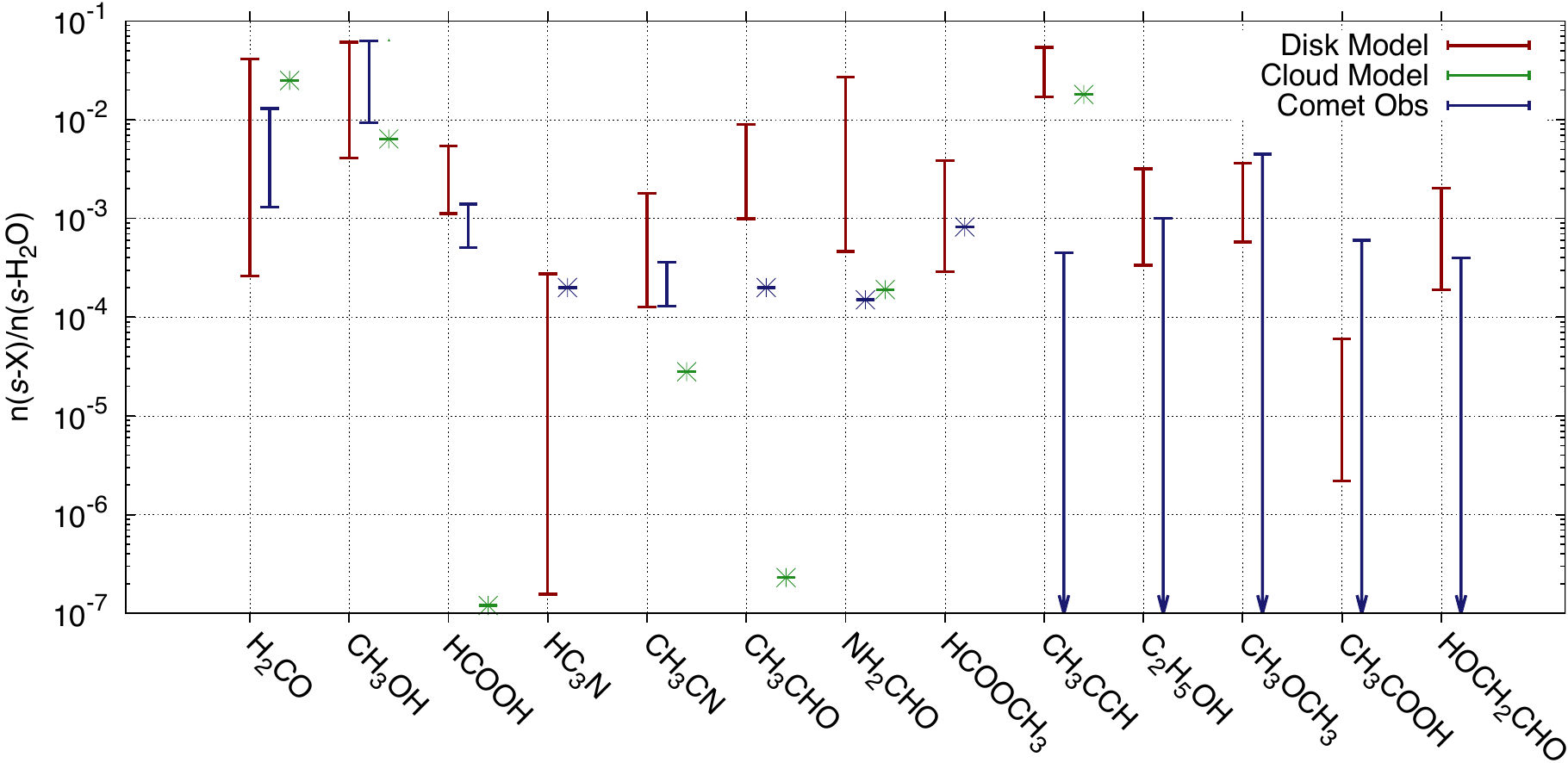}
\caption{\label{fig:walsh}Complex organics can be formed by chemistry
  in the ice mantles of dust grains in a protostellar disk.  This
  figure shows the predicted range of abundances of some of these
  molecules relative to water ice (red lines) compared to those
  derived from cometary comae \citep[blue
  lines;][]{bm04,crov06,crov04}.  The green asterisks are the results
  from the initial molecular cloud model used as the input for the
  disk model.  From \cite{walsh14}.}
\end{figure}

Of particular interest to the chemistry of comets is the work of
\cite{walsh14} who modeled the possible formation of complex organics through ice
chemistry in protostellar disks. Figure~\ref{fig:walsh} shows the
range of abundances they find in ices located outside of 20 AU in a
solar-type disk.  The large molecules form by sequential addition of
atoms to existing mantle molecules, e.g., ethylene glycol can form by
adding hydrogen, oxygen and carbon atoms to CO:
\begin{align*}
\hbox{CO} & \xrightarrow{\hbox{H}}  \hbox{HCO}
\xrightarrow{\hbox{C}} \hbox{HC$_2$O}
\xrightarrow{\hbox{O}} \hbox{OCCHO}  
\\ 
\hbox{OCCHO} & \xrightarrow{\hbox{H}}  \hbox{CHOCHO} 
\xrightarrow{\hbox{2H}} \hbox{HOCH$_2$CHO} \xrightarrow{\hbox{2H}} \hbox{(HOCH$_2$)$_2$}
\end{align*}
\citep[][see Figure~\ref{fig:stern}]{cr08}.  The models of Walsh et al. suggest that chemistry in disks will 
greatly increase the abundance of these large
molecules relative to what can be created in the parent cloud.  The
predicted abundances relative to H$_2$O are consistent in many cases with
observations of comets \citep{bm04,crov06,crov04}.  If these molecules
can be desorbed they will also be present in the disk gas, and should
be detectable by ALMA providing a test of the model.

\subsubsection{\label{sec:turb}The effects of dynamics on the chemistry}
There are strong links between the disk chemistry and its dynamics.
The temperature, density and radiation field in a disk control the
chemistry, but the chemistry (and grain evolution) in turn controls
the dynamics of the disk through the ionization structure.  Models
have found that better agreement with observations of the molecular
layer in the outer disk can be obtained if vertical mixing is
included, e.g., \cite{willacy06,semenov06}, and more sophisticated
models have now considered 2-D mixing
\citep[e.g.,][]{heinzeller11,sw11,albertsson14}.  Mixing smooths out
chemical abundance gradients but also brings together species that
might not otherwise exist in the same location, e.g., in the inner
disk, vertical mixing combines OH and H$_2$ in a region where the
temperature is high enough for the neutral-neutral reaction, \mbox{OH +
H$_2$ = H$_2$O + H} to occur \citep{heinzeller11}.  As a result the
abundance of H$_2$O increases.

\begin{figure}[!tb]
\center
\includegraphics[width=0.8\linewidth]{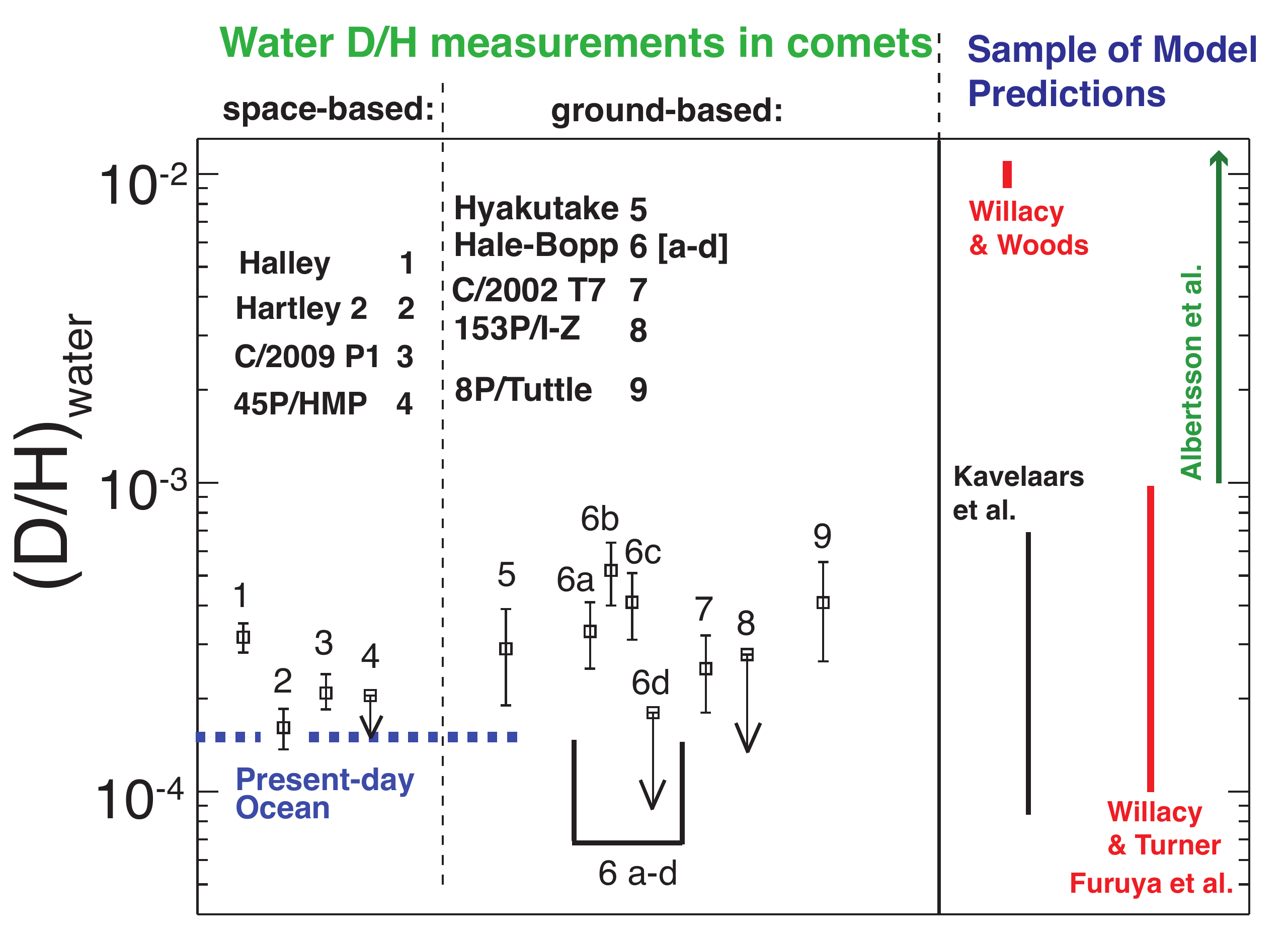}
\caption{\label{fig:comets}\label{fig:dh}Current disk models can only
  account for the observed D/H ratios in comets if they include
  mixing.  The non-mixing models of \cite{willacy_woods09} and
  T. Millar (private communication) predict D/H ratios that are
  consistently higher than the observations.  \cite{albertsson11} treat
  the evolution of the molecular cloud to the hot core stage only and
  predict (D/H)$_{water}$ $>$ 10$^{-3}$.  Turbulent mixing is included
  in the disk models of \cite{kavelaars11} (radial mixing) and
  \cite{willacy13} (vertical mixing at 5 AU) and these produce
  D/H$_{water}$ in agreement with the observations.  \cite{furuya13}
  predict similar D/H ratios to \cite{willacy13}. References: (1)
  \cite{eberhardt95}, (2) \cite{har}, (3) \cite{bm12}, (4) \cite{lis13},
 (5)  \cite{bm98} (6a-d) \cite{meier98} and \cite{crov04}, (7)
  \cite{hut08}, (8) \cite{biver06}, (9) \cite{villa09}.  Adapted from
  \cite{boncho14}.}
\end{figure}

In addition to affecting molecular abundances, mixing also changes
isotopic ratios. One problem with the non-mixing models is the high
deuteration levels predicted in the comet formation zone, e.g.,
\cite{willacy_woods09}.  These reflect the high levels set in the
parent molecular cloud, whereas those detected in comets tend to be
somewhat lower.  As D/H ratios track the thermal history of the
material this suggests that at least some of the cometary ices formed
at warmer temperatures than are found in the interstellar medium (ISM),
e.g., \cite{mo99}.  Figure~\ref{fig:comets} compares the observed
molecular deuteration in cometary water with the predictions of several
models.  The non-mixing models uniformly over-predict HDO/H$_2$O, but
the addition of mixing (either radial or vertical) brings the ratio
into closer agreement with the observations \citep{boncho14}.

With vertical mixing this reduction in the deuteration of water is
achieved by the destruction and reformation of ice mantles. When dust
grains are mixed vertically they can reach the surface layers where
high temperatures and/or UV fields desorb their ices.  The water
molecules are dissociated and the resulting oxygen, hydrogen and
deuterium atoms travel back down towards the midplane into more
shielded regions where they reform molecules.  The molecular
deuteration depends on the formation temperature.  Since the disk is
warmer than the 10 K parent molecular cloud core, the reformed water
ice has a lower D/H ratio.  Hence vertical mixing can reduce the
initial interstellar D/H ratio in water ice $\sim$ 10$^{-2}$) to the
level detected in comets ($\sim$ 10$^{-4}$)
\citep[Figure~\ref{fig:comets} and see also][]{furuya13}.
  \cite{willacy13} and \cite{furuya13} also predict radial
variations in the deuteration of water.   A radial gradient in
HDO/H$_2$O is also found in the (radial) mixing models of
\cite{kavelaars11}.

\subsubsection{Grain evolution}

\begin{figure}[!htb]
\includegraphics[width=\linewidth]{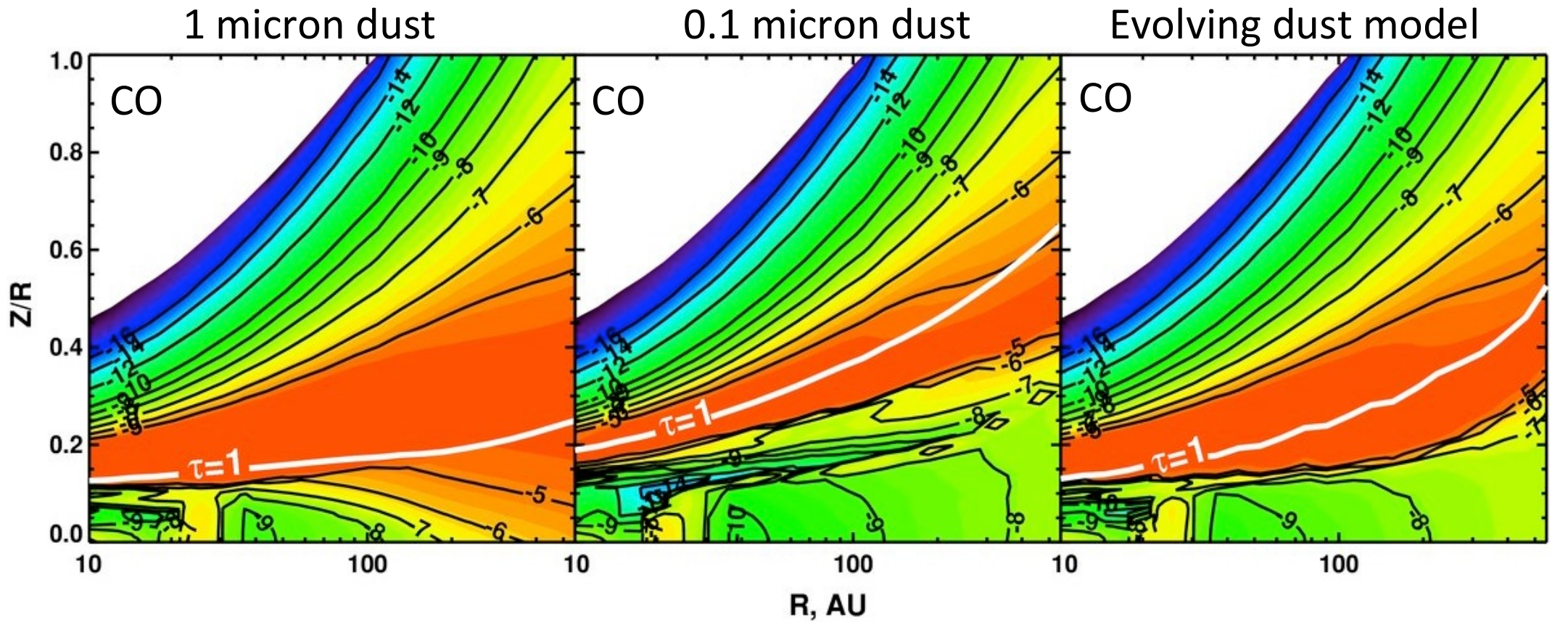}
\caption{\label{fig:graingrowth}The CO abundance distribution in a
  protostellar disk, showing the effects of grain growth.
The highest abundances are shown in red.  Grain
  growth reduces the opacity in the surface layers, allowing UV to
  penetrate further and pushing the molecular layer closer to the
  midplane.  With 1 micron dust (left) the molecular layer is much
  thicker, and reaches closer to the midplane than in the model with
  0.1 micron dust (center).  On the right are the results from an
  evolving dust model where the grains are allowed to both grow and
  sediment. In this case the molecular layer is thicker than in the
  small grain model, and it moves closer to the midplane.  From
  \cite{vasyunin11}. \textcopyright AAS. Reproduced with permission.}
\end{figure}
 Grain-grain collisions in protostellar disks lead to efficient grain
growth and ultimately result in the formation of planetesimals.  
  As the grains grow they will decouple from
the gas motions and sink towards the midplane.  This removes grains
from the upper layers of the disk, reducing the opacity and allowing
UV to penetrate deeper, which increases photodissociation and heating of the newly
irradiated layers, and moves the molecular layer closer to the midplane.
An additional effect of grain growth is a reduction in the freezeout rate, which
depends on the surface area of the grains.
Figure~\ref{fig:graingrowth} shows the effect on the chemistry of CO
in the models of \cite{vasyunin11} 
\citep[see also][]{an06,walsh14}.
Vasyunin et al. found that CO, CN and SO, as well as the ratio of
C$_2$H$_2$/HCN, 
should be good tracers of the grain growth.

\subsection{\label{sec:snowlines}Gas-dust chemistry and icelines in
  protoplanetary disks}

  Icelines in the early nebula are defined as radii at which specific
  molecules undergo a shift in abundance from the gas phase to the
  condensed phase. There are three of specific interest related to
  H$_{2}$O, CO$_2$ and CO and these occur at radii where the
  temperature is approximately 150 K, 47 K and 20 K respectively
  \citep{oberg11}.
Recent observations by ALMA have allowed icelines to be identified in
protoplanetary disks.  For example, in HD 163926
\citep{qi11,dgm13,mathews13}.  
   The icelines associated with H$_{2}$O, CO$_2$, and CO
  are regions of dramatic chemical change and these could be reflected
  in the molecular composition of the gas and ice that is transported
  and mixed, both vertically and radially, in the evolving disk.
  
  The chemical signature of such an environment may be imprinted in
  cometary and meteoritic composition if the Sun formed in a stellar
  cluster containing at least one massive star \citep{adams10,mumma}.
   Nearby OB stars strongly affect the positions of
  the CO$_2$ and CO icelines and the chemical structure of the disk
  \citep{walsh13}.
  The CO iceline occurs at tens of AU from the
  central protostar and is most accessible to observations at
  millimeter and submillimeter wavelengths.  The physical conditions
  here most closely resemble those in the densest molecular cloud
  cores and so the related gas-dust chemistry could also be similar
  \citep{hs13}.  This chemistry can lead to formation of
  many complex organic molecules on cold dust grains, as well as
  potentially large isotopic fractionation in both gas phase and
  grain-surface reactions.
  
\begin{figure}[!tb]
  \begin{center}
\includegraphics[angle=0,width=4.0in]{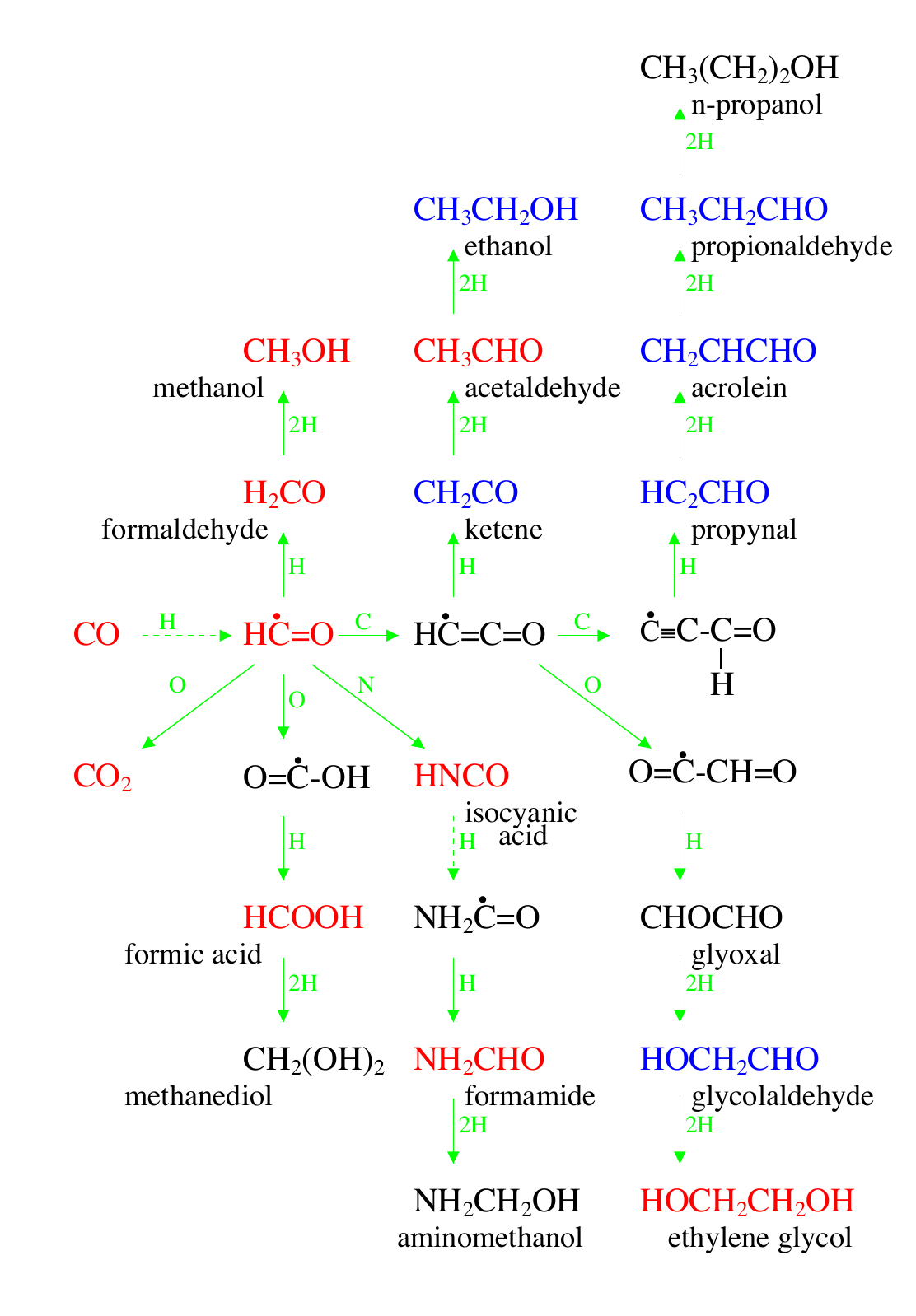}  
\caption{Grain surface reaction scheme forming organics via atom
  addition reactions to CO on cold dust. Molecules in red are detected
  both in comets and in interstellar/protostellar sources, whereas
  those in blue are found only in the latter. Species in black are not
  yet detected in space.  The addition of deuterium atoms rather than
  hydrogen atoms to grain species can result in high levels of
  molecular deuteration. Adapted from \cite{cr08}.}
\label{fig:stern}
  \end{center}
  \end{figure}

  For example, outside the CO iceline, CO-rich ices on dust grains
  can undergo (tunneling) addition reactions with H (and D) atoms
  accreted from the gas \citep{tielens83,charnley97}.
  Atoms of O, C and N will also accrete and
  these processes can lead to rich organic chemistry
  \citep[Figure~\ref{fig:stern};][]{charnley97,cr08,hvd09}.  Many of
  the proposed hydrogenation sequences have now been demonstrated in
  surface chemistry experiments \citep{wk08,theule13}
   and such an ice chemistry explains the presence of the
  distinctive and inter-related suite of organic molecules detected
  around massive and low-mass protostars (in hot cores and hot
  corinos, respectively), where the ice mantles have been evaporated
  into the hot gas \citep{hvd09}.
   Figure~\ref{fig:stern} indicates that many of the relevant
  organic molecules could also be present in comets and that, if
  correct, ketene, ethanol, propynal, acrolein, propionaldehyde and
  propanol should be searched for in suitably bright apparitions.
  Several of the organic compounds formed in the scheme of
  Figure~\ref{fig:stern} may be detectable in protoplanetary disks
  with ALMA.  Vertical transport of ice-covered dust to the less
  well-shielded layers of the disk, as well as (inward) radial
  transport, can lead to these organic compounds being released by
  evaporation and/or photodesorption. This is probably the origin of
  the H$_{2}$CO detected in HD 163296, which lies outside of the CO
  iceline at about 160 AU \citep{qi13b}.

\subsection{\label{sec:h2o}Two distinct reservoirs of water ice in the protosolar nebula}

\begin{figure}
\center
\includegraphics[width=0.9\textwidth]{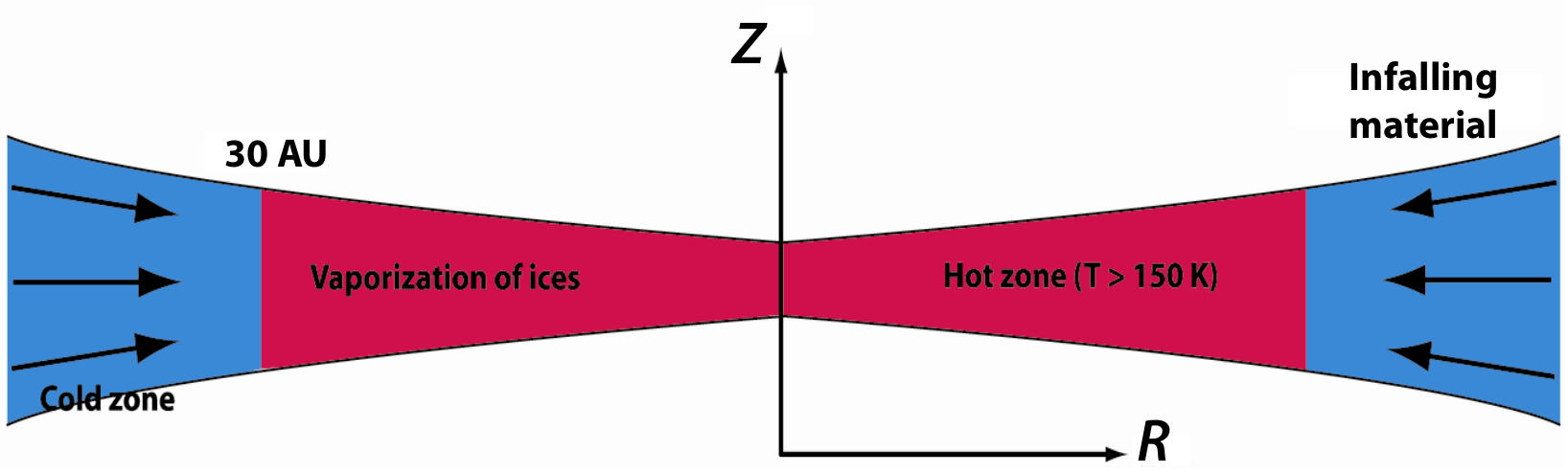}
\caption{Two reservoirs of ices in the PSN. A first reservoir (cold
  zone), located at distances higher than $\sim$30 AU, is constituted
  from amorphous ices coming from ISM. The second reservoir (hot
  zone), located within the $\sim$30 AU, is made from volatiles
  initially in the form of amorphous ices that were transported from
  ISM towards the inner and hot part of the disk. When reaching
  regions with temperatures higher than $\sim$150 K, these ices
  vaporized. During the cooling of the disk, volatiles located in the
  inner 30 AU condensed again but in crystalline forms, including both
  pure crystalline ices and clathrates (see text).} 
\label{plot_neb}   
\end{figure}

Formation scenarios of the protosolar nebula (PSN) invoke two main
reservoirs of ices present in the disk midplane and that took part in
the production of icy planetesimals (see Figure~\ref{plot_neb}). The
first reservoir, located in the inner region of the PSN, contains ices
(dominated by H$_2$O, CO, CO$_2$, CH$_4$, N$_2$ and NH$_3$)
originating from the ISM, which, due to their near solar vicinity,
were initially vaporized. The ice vaporization distance never exceeds
$\sim$ 30 AU from the Sun, depending on the total source luminosity and
characteristics of the collapsing cloud \citep{Chick97}. With time,
the decrease of temperature and pressure conditions allowed the water
in this reservoir to condense at $\sim$150 K (at typical nebula
pressure conditions) in the form of microscopic crystalline ice
\citep{Kouchi}, leaving negligible water in the vapor phase to
condense at lower temperatures where amorphous ice would be
expected. It is postulated that a substantial fraction of the volatile
species were trapped as clathrates during this condensation phase as
long as free water ice was available within 30 AU in the outer solar
nebula \citep{mou00} and there was enough time to overcome the
kinetics of clathration. On the other hand, the remaining volatiles
that were not enclathrated (due to the lack of available water ice or
a low kinetics of clathration) probably formed pure condensates at
lower temperatures in this part of the nebula
\citep{m2012a,m2012}. The other reservoir, located at larger
heliocentric distances, is composed of ices originating from the ISM
that did not vaporize when entering into the disk. In this reservoir,
water ice was essentially in the amorphous form and the other
volatiles remained trapped in the amorphous matrix
\citep{Owen,Notesco}.  

Interestingly, it has been shown recently that amorphous ice could
form from photodesorption and freeze-out of water molecules near the
surface layers of the PSN \citep{Ciesla}. In these conditions, the
transport of the icy grains throughout the outer solar nebula would
lead crystalline ice to be lost and reformed as amorphous ice. In
turn, amorphous ice generated on the disk's surface could be
transformed to crystalline as it migrates through the protoplanetary
disk \citep{Ciesla}. This cycling scenario needs to be further
  investigated  if one wants to assess its role in the formation or
  destruction of the two original icy reservoirs. Also, the different
  transport mechanisms must be simultaneously taken into account at
  the different stages of the disk's evolution to depict the abundance
  of water ice in the two reservoirs as a function of the heliocentric
  distance. For example, as described in Sec. 2.3.2, it has been shown
  that vertical mixing can decrease the abundance of water inside the
  $\sim$15 AU of the disk. In contrast, it has also been proposed that
  the outward vapor diffusion induced by the presence of the iceline
  with its subsequent condensation in a narrow location (with a width
  around 0.5 AU) would deplete the region inside the iceline of vapor
  down to subsolar values, and increase substantially the ice
  abundance at the iceline location by factors larger than 10 times
  the protosolar value \citep{mad}. In addition, recent works
  depicting the chemical evolution of both solids and gases from the
  pre-stellar core to the protostar and circumstellar disk phases
  \citep{VISSER09,VISSER11,HAR13} should be taken into account to
  update the representation of the two water ice reservoirs present in
  the PSN. In these cases, midplane ices may have experienced multiple
  desorption events during the evolution of the disk. Nevertheless,
if one follows this classical picture, icy solids that formed at
heliocentric distances less than $\sim$30 AU mainly agglomerated from
a mixture of clathrates and pure condensates, whose ratio depends on
the amount of available crystalline water and its clathration
efficiency \citep{mou09b}. In contrast, solids produced at higher
heliocentric distances (i.e., in the cold outer part of the solar
nebula) were formed from primordial amorphous ice originating from the
ISM. Therefore, depending where the outer solar system bodies formed,
clathrates may have been agglomerated in comets
\citep{mar10,mar11,mar12}, and in the building blocks of the giant
planets \citep{gaut,ali05a,ali05b,mou09b,m2012a} and in their
surrounding satellite systems
\citep{lunine,mou04,MG04,mou09a}. Regardless the possible presence of
clathrates, there is today indirect evidence that comets formed from
crystalline water ice because their deuterium-to-hydrogen ratios
measured in H$_2$O are substantially lower ($\sim$6--12 times the
protosolar value) than the most deuterium-rich primitive meteorite
($\sim$35 times the protosolar value), whose level of deuteration is
expected to be close to the one found in the ice infalling from the
presolar cloud \citep{kavelaars11,har,cec}. This difference can only been
explained by the isotopic exchange between the deuterium-rich water
and molecular hydrogen in the gas phase \citep{kavelaars11,har,cec}. Hence, in
order to enable this reaction, amorphous ice entering the disk from
ISM must have been vaporized prior recondensing again in crystalline
form, otherwise the isotopic exchange did not occur.

\section{\label{sec:fractionation}Fractionation in protostellar disks}
\subsection{\label{sec:deut}The deuterium fractionation: The link between
  comets and the earliest phases of the solar system and interstellar
  chemistry}

  The origin of the volatile ices found in comets is a central issue
  for understanding the formation and early evolution of the 
  solar system. There are three possibilities. First, one view is that
  {\it all} the ices are comprised of pristine interstellar molecules
 \citep[e.g.][]{irvine96}. There is indeed a strong similarity with
  the many known interstellar molecules believed to form on
  interstellar grains, although there are differences in the relative
  abundances of the volatile ices \citep{mumma}. Until
  recently this idea was widely considered to be untenable (see
  Section~\ref{sec:cleeves} and
  Cleeves et al. 2014a).  Second, at the other extreme, it has been
  advocated that any interstellar molecules experienced total
  obliteration and `chemical reset' in the disk \citep{pont14a}. 
In this case all the volatile molecules observed in
  comets must have been formed by nebular chemistry.
  
Comets and meteorites contain both high-temperature processed
materials, such as crystalline silicates, and matter that apparently
formed at very low temperatures, as indicated by the isotopic
enrichments measured in D and $^{15}$N (Mumma \& Charnley 2011). These
characteristics are most naturally explained by the large-scale
turbulent transport of dust and gas from different nebular chemical
environments into the comet-formation zones \citep{cc06}.
The recognition that non-equilibrium models of nebular chemistry are
essential, particularly in the comet-forming regions, has motivated
many detailed studies of protoplanetary disk chemistry, as recently
reviewed by \cite{hs13} (see also Section~\ref{sec:turb}).  Turbulence
leads to an outward radial diffusion of material from the hot inner
nebula and an inwards advection of material from the cooler outer
nebula. It also leads to vertical mixing and thus to midplane material
experiencing the enhanced UV and X-ray fluxes of the upper disk layers
(Glassgold et al. 1997).  Hence, in this third partially-mixed
scenario some specific volatile molecules and isotopic ratios could be
remnant interstellar matter, or derive from warm and cold regions of
the nebula \citep[as reviewed by ][]{pont14}.  Deciding between such molecular origins has been
problematic. 
 
\subsubsection{\label{sec:moldeut}The molecular deuteration process}
Molecular deuteration provides a potential clue to the
history of cometary volatiles.
 Although deuterium in the Universe is a tiny fraction with respect to
the hydrogen (D/H=[2.535$\pm$0.05]$\times10^{-5}$: Pettini \& Cooke
2012), its abundance in interstellar molecules has an enormous
diagnostic power, because it may be much larger that the elemental D/H
ratio. This generally occurs in cold ($\leq$ 30--50 K) environments and
in trace species, namely molecules with abundances smaller than
$\sim10^{-5}$. 

{\it The first key point to keep in mind is that, in cold gas, the
  major deuterium reservoir is HD and only a small fraction of
  deuterium is locked in other species.} An exception to this rule is
represented by the water, whose abundance can reach $\sim10^{-4}$ and,
consequently, water might trap a substantial fraction of deuterium.

\begin{figure}[!tb]
\center
\includegraphics[width=0.9\textwidth]{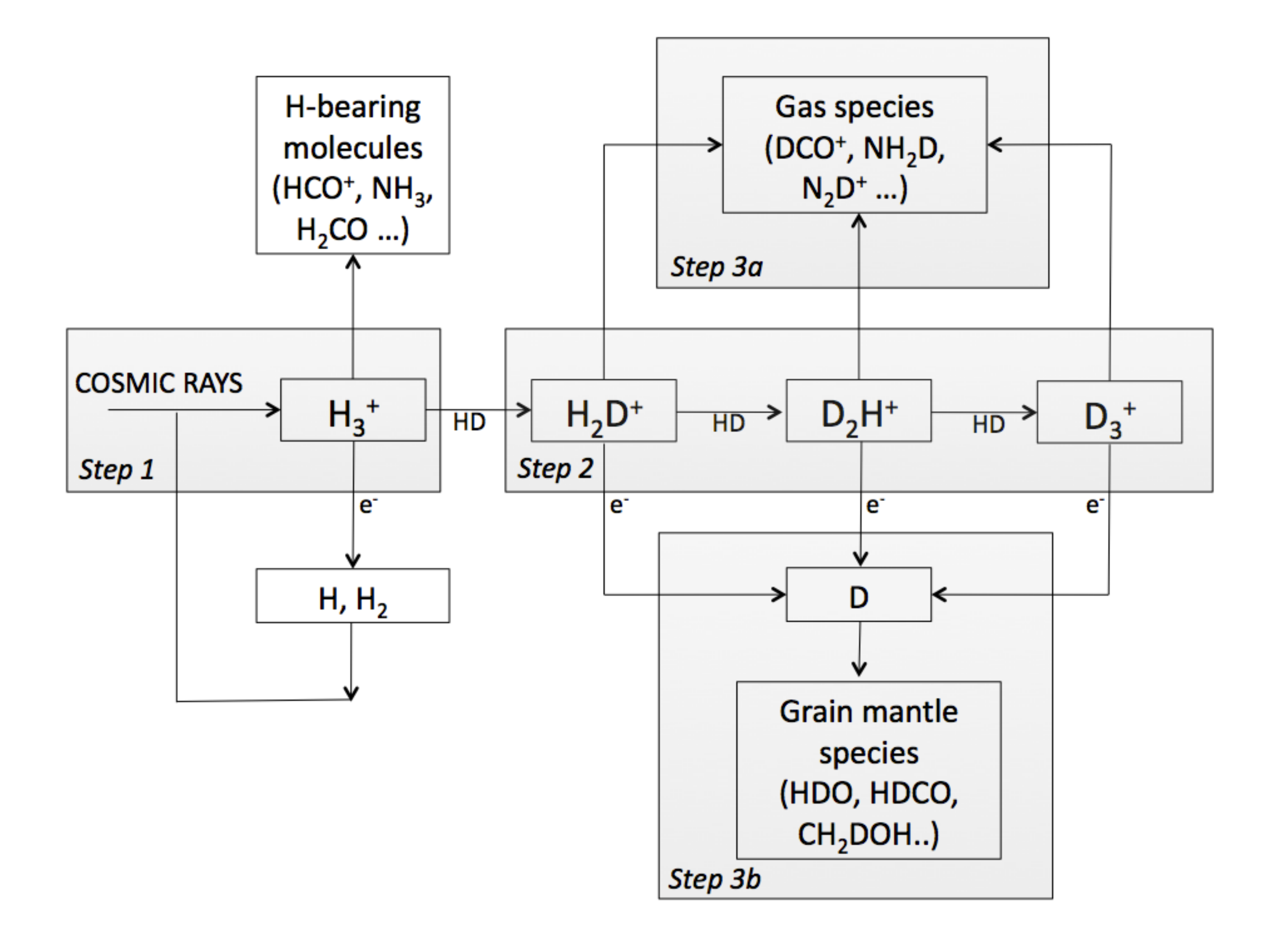}
\caption{\label{schema_deuteration} Schema of the molecular
  deuteration process in cold gas \citep{cec}.
  Molecular deuterium occurs through three basic steps: 1)
  formation of H$_3^+$ ions from the interaction of cosmic rays with H
  and H$_2$; 2) formation of H$_2$D$^+$ (HD$_2^+$ and D$_3^+$) from
  the reaction of H$_3^+$ (H$_2$D$^+$ and HD$_2^+$) with HD; 3)
  formation of other D-bearing molecules from reactions with
  H$_2$D$^+$ (HD$_2^+$ and D$_3^+$) in the gas phase (Step 3a) and on
  the grain mantles (Step 3b).  }
\end{figure}

{\it The second key point is that the deuterium atoms are
  ``extracted'' from the HD molecules because the cosmic rays that hit H
  and H$_2$ create H$_3^+$ ions, which in turn react with HD and form
  H$_2$D$^+$ ions.}
        \begin{equation} {\rm H_3^+ ~+~ HD } ~\rightleftharpoons~ {\rm
            H_2D^+ ~+~ H_2~+~220K} 
\label{eq:h2d+}
\end{equation}
(see also Figure~\ref{fig:stern}). The reverse reaction has a small energy barrier,
which cannot be overcome at low ($\leq$ 20--30 K) temperatures,
so that the H$_2$D$^+$/H$_3^+$ abundance ratio becomes larger than the
D/H elemental ratio.  The kinetics of Reaction~\ref{eq:h2d+} at low
temperatures is also strongly influenced by the spin state of the
reacting hydrogen molecules, i.e., whether they are predominantly in
the $\it ortho$ or $\it para$ states.  The higher internal energy of
$\it ortho$-H$_2$ means that the reverse process can be driven
efficiently if the H$_2$ $\it ortho/para$ ratio is large.  Thus, $\it
ortho$-H$_2$ acts as a ``poison'' for interstellar deuteration which
requires that most of the H$_2$ be converted to $\it para$-H$_2$
\citep[see][]{pagani11}.  Reaction~\ref{eq:h2d+} is the first step towards the deuteration of
the trace species, as H$_2$D$^+$ reacts with all neutral molecules and
passes, once every three times, the D atom to the product species.
This is schematically shown in Figure~\ref{schema_deuteration}. A
similar process can also occur through the CH$_3^+$ ions at slightly
higher temperatures.

In practice, therefore, a molecular deuteration enhanced with respect
to the D/H elemental abundance is a thermometer of the gas
temperature.

\begin{figure}[!tb]
\center
\includegraphics[width=0.9\textwidth]{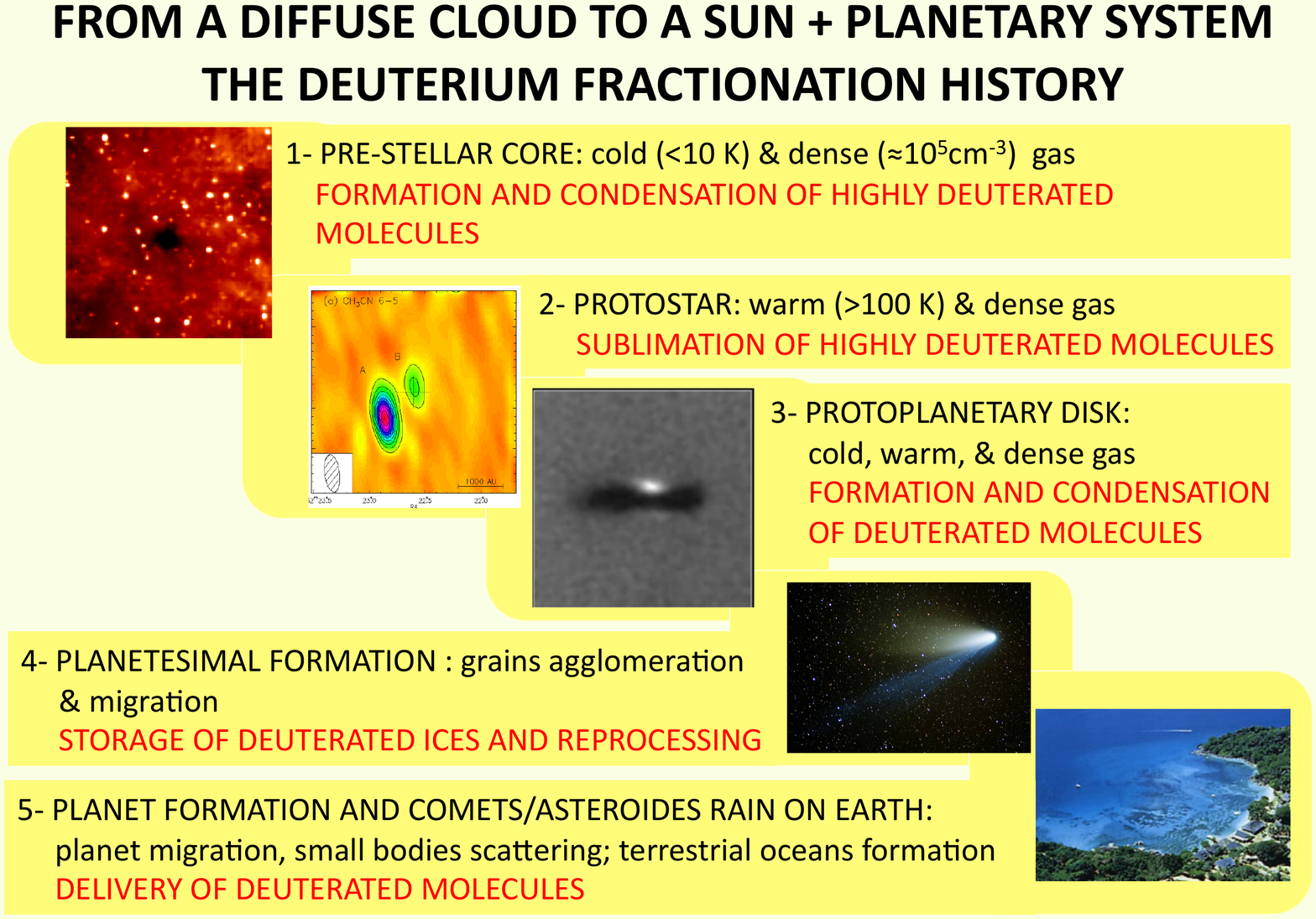}
\caption{\label{schema_evolution} Schematic summary of the different
  phases of the formation of a solar-type star and planets, 
  highlighting the formation and fate of deuterated molecules (adapted
  from \cite{cec}. Likely, the same process gave birth
  to the solar system.}
\end{figure}

{\it The third and final key point to keep in mind is that neutral
  molecules can freeze-out into the grain mantles and, in this way,
  they can be transmitted through the different stages of the star and
  planet formation.}  A sketch of this overall process is shown in
Figure~\ref{schema_evolution}.
Therefore, ice species formed during the very first phases might be
conserved and ``reappear'' in the gas at much later stages: for
example when comets and planets are formed. Conversely, the molecular
deuteration in comets and planets can help us to understand in what
epoch the relevant species was formed, at least in principle. In
this sense, molecular deuteration can be thought of as an Ariadne's
thread which links the various phases of solar system evolution.  For
a more detailed review see \cite{cec}.

  \subsubsection{\label{sec:cleeves}Deuteration in the solar
   nebula}

\begin{figure}[!htb]
\center
\includegraphics[width=\linewidth]{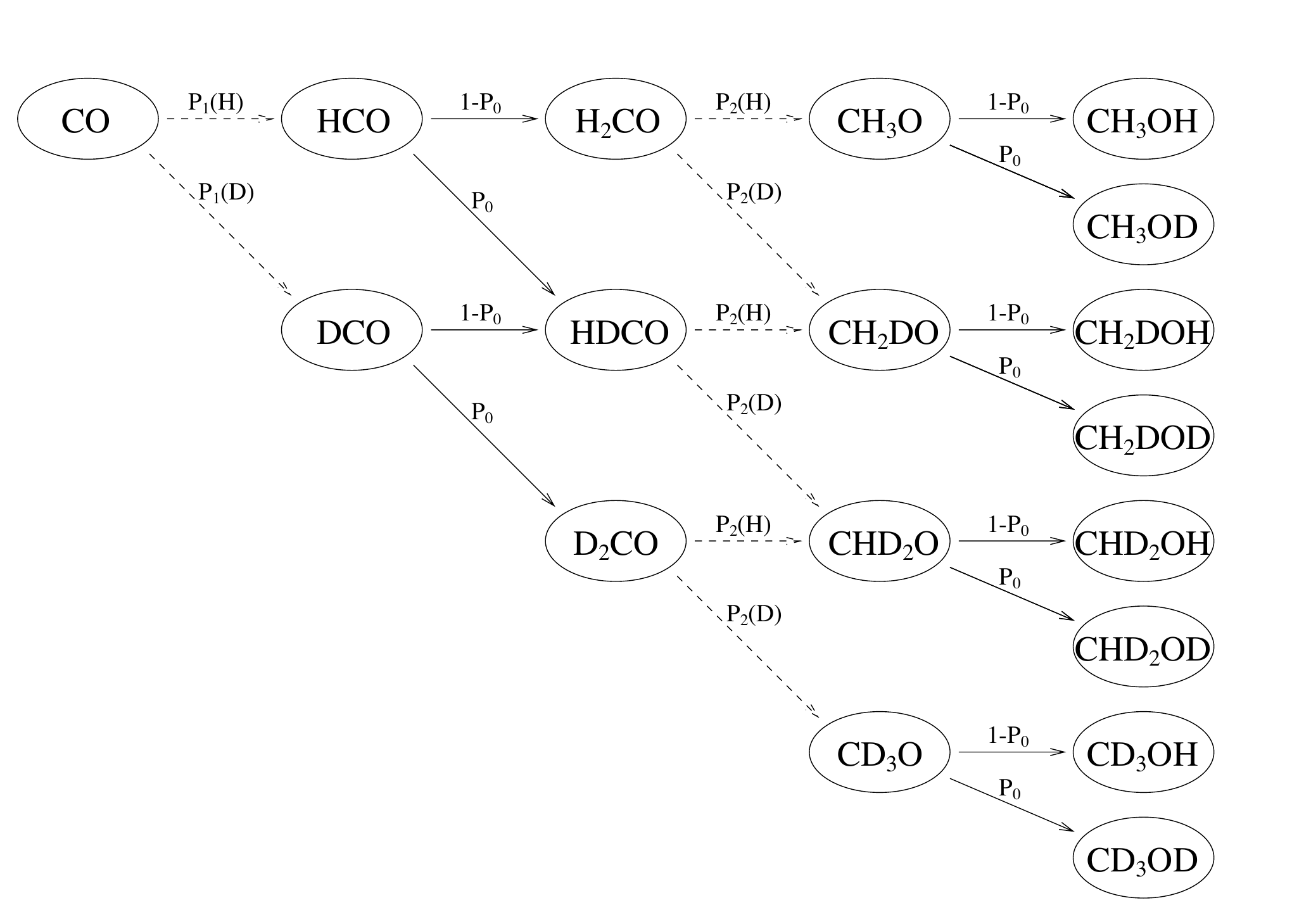}
\caption{\label{fig:steve_deut}Grain surface deuteration pathways
  associated with methanol formation from hydrogenation of CO ice.
  The probabilities for each reaction pathway, P$_i$, assume a statistical steady state and are
  related to the atomic D/H ratio of the accreting gas and the energy
  barrier for addition of an H or D atom \citep[from][]{ctr97}.}
\end{figure}

 Deuteration in disks is driven mainly via the reaction of H$_2$D$^+$
 ions (Reaction~\ref{eq:h2d+}).  At low temperatures the reverse
 process is inhibited and this leads to enhanced deuteration in
 gaseous molecules through ion-molecule isotope exchange
 reactions. Elevated atomic D/H ratios also pertain and lead to
 significant deuteration in grain-surface reactions, such as those
 depicted in Figure~\ref{fig:steve_deut} and in the surface reduction
 of accreted O atoms to form HDO.  Condensation of CO leads to
 enhanced abundances of gaseous H$_{2}$D$^+$ and to significant
 deuterium fractionation beyond the CO iceline, as traced by DCO$^+$
 \citep{mathews13}.  Enhanced D/H ratios in simple molecules, such as
 water, ammonia and methane, can be produced both in ion-molecule
 reactions and grain-surface additions of H and D to accreted atoms
 \citep{cec}.  At the CO iceline the grain-surface chemistry depicted
 in Figure~\ref{fig:steve_deut} can lead to very high D/H ratios and
 multiply deuterated molecules \citep{ctr97,cec}.  Observations show
 that simple deuterated species like DCN and by DCO$^+$ are present in
 disks \citep{qi08} and we would expect that more complex fractionated
 organics, e.g., HDCO, D$_2$CO, CH$_{3}$OD and CH$_{2}$DOH, to also be
 present. Between the CO and CO$_2$ icelines (see
 Section~\ref{sec:snowlines}), large deuterium enrichments can persist
 in organic molecules since ion-molecule deuteration involving
 C$_2$HD$^+$ and CH$_{2}$D$^+$ can persist up to higher temperatures
 ($\sim 70$ K and $\sim 50$ K) than H$_{2}$D$^+$ ($\sim 30$K)
 \citep{millar89}. It is possible that the D/H ratios of several
 organic molecules can become re-fractionated even in the warm inner
 nebula, prior to the condensation of acetylene and methane, both of
 which will occur outside the CO$_2$ iceline.  This means that
 deuteration can still occur in cold gas that is mixed radially
 inwards, and be initiated in warmer gas coming from the inner nebula
 \citep{willacy07,albertsson14}.

However, the assumption of interstellar rates of cosmic-ray driven
ionization ($\zeta$) in the cold regions of the outer midplane has
recently been called into question. \cite{cleeves13} showed that the
effects of either deflection by stellar winds or magnetic field
topology (through the magnetic mirror effect) could significantly
attenuate the penetration of cosmic rays into magnetized
protoplanetary disks, and substantially reduce $\zeta$ by several
orders of magnitude at the midplane.  In this picture the only sources
of ionization close to the midplane are scattered X-rays and the decay
of short-lived radionuclides.  In fact, \cite{ds94} had previously
pointed out that inhomogenous random magnetic fields could also
significantly inhibit the transport of cosmic rays throughout the disk
and lead to very low values of $\zeta$.
  
\cite{cleeves14a} developed this scenario further by pointing out
that effectively extinguishing cosmic ray ionization from the nebular
midplane would greatly reduce the efficacy of Reaction~\ref{eq:h2d+} and
consequently the generation of enhanced molecular D/H ratios by
nebular ion-molecule chemistry.  They demonstrated
that if the enhanced interstellar HDO/${\rm H_2O }$ ratios initially
available were lost, then nebular chemistry could not regenerate those
currently measured in the solar system, concluding that as much as
50\% of the Earth's oceans and perhaps {\it all} cometary water is of
interstellar origin.  Low values of $\zeta$ will also reduce the
atomic D/H ratio in the disk midplane and so reduce the D/H ratios and
the level of multiple deuteration in the molecules formed in the
scheme of Figure~\ref{fig:steve_deut}.  This may account for the upper limits on HDCO,
CH$_2$DOH and CH$_3$DOD in comet Hale-Bopp \citep{crov04}
relative to the large observed interstellar D/H ratios measured in
these molecules, which could have been lost or modified upon
incorporation into the nebula. Although these molecules could be
partially reformed in the cold outer nebula, this process, and any
related deuteration, may be less efficient than in molecular clouds.
Cometary CH$_3$OH/ H$_2$O ratios are typically $\sim 1-4 \%$, whereas
$\sim 5-15 \%$ are typical for the ISM, and can be as high as 40\%
\citep{mumma}.  \cite{cleeves14b} also modeled the
general protoplanetary disk chemistry for a variety of assumed $\zeta$
values and identified molecular ions that may eventually allow the
ionization structure to be determined.  If these spatial distributions
are observable with ALMA they could provide valuable information on
the actual role of cosmic rays in nebular chemistry.

\subsubsection{The message from molecular deuteration}
\begin{figure}[!htb]
\center
\includegraphics[width=0.9\textwidth]{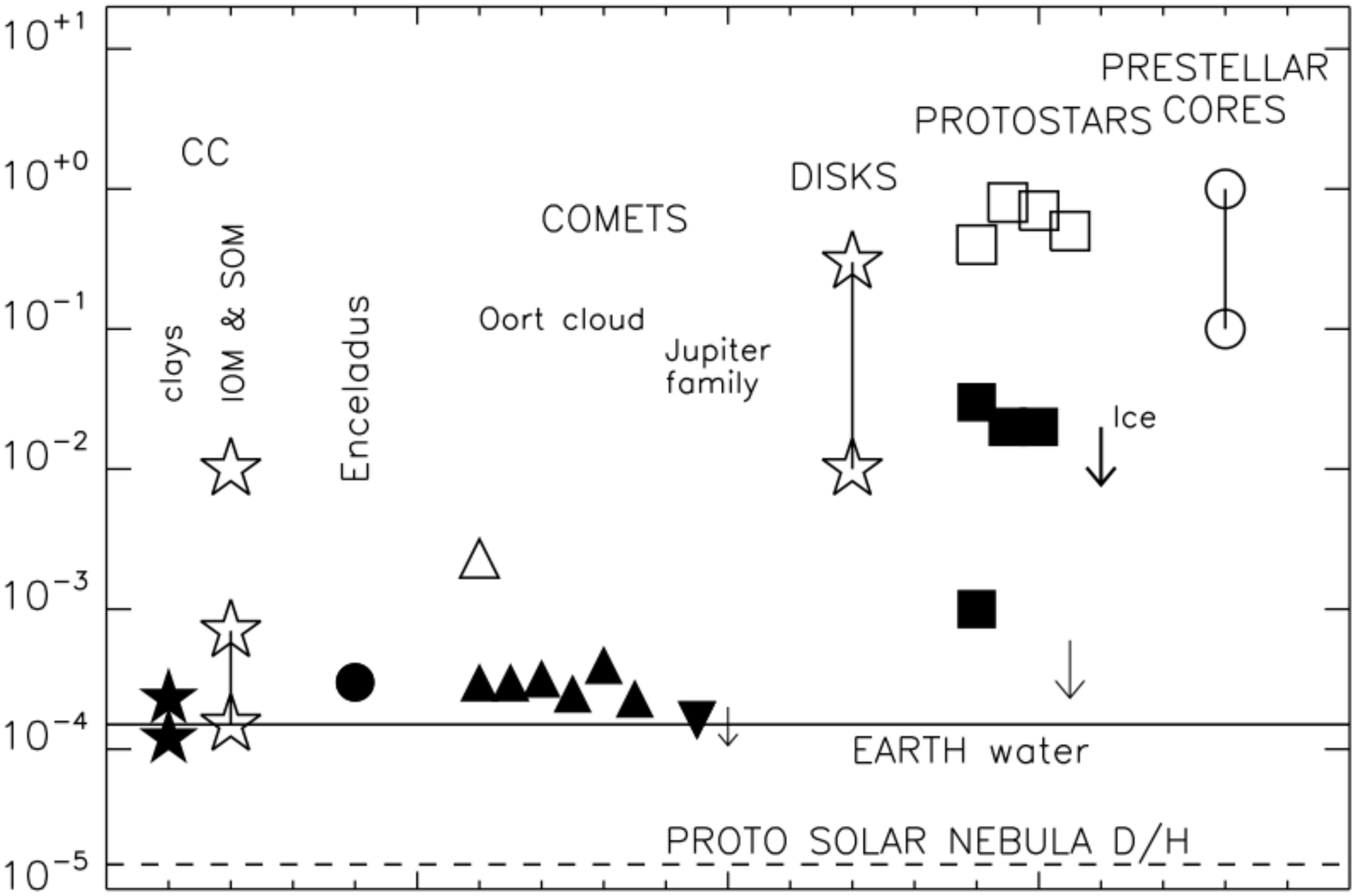}
\caption{\label{deuteration_summary} D/H ratio in molecules observed
  in solar system objects and in objects which are believed to
  eventually form a solar-like system. The filled symbols refer to
  water measures whereas open symbols refer to organic matter. The
  different objects are marked (note: CC means Carbonaceous
  Chondrites, IOM Insoluble Organic Matter, and SOM Soluble Organic
  Matter). References to the various measures can be found in
\cite{cec}.}
\end{figure}

Figure~\ref{deuteration_summary} shows the molecular deuteration in the
different objects of the solar system and in objects which are
believed to eventually form a solar-like system.
It gives us the following major messages:\\
\begin{itemize}
\item All reported solar system objects, including Earth, possess a
  molecular deuteration larger, by more than a factor 10, than the D/H
  elemental ratio of the environment where the Sun was born, and
  that is represented by the line marked with ``Proto Solar
  Nebula''. This implies that the process of molecular deuteration was
  at work and it was at least partially inherited from the earlier
  phases of the solar system formation era.
\item There is a general trend, with the objects that formed earlier
  being also the ones with the largest molecular deuteration, in
  agreement with the scenario that the process started in very cold
  conditions. This implies that there was only a partial transmission
  between the subsequent steps of
  Figure~\ref{schema_evolution}. Besides, earlier objects in the plot
  also correspond to material whose distance from the central heating
  source is larger.  Therefore, the trend also tells us that there was
  not a general and substantial remixing of the different regions
  (outer and inner, where molecular deuteration is different because
  of the different temperatures). Exactly how much material was passed
  through each step or remixed remains, however, unanswered.
\item The organic matter is systematically more enriched in deuterium
  atoms than the water. This is true in protostellar objects, comets
  and meteorites. However, while this result is robust for the
  protostellar objects and meteorites, where several measurements of
  the deuteration have been made, it is less certain for the
  comets, where so far the only organic species for which 
  molecular deuteration has been measured is HCN.  
  The reason for this dichotomy in the molecular
  deuteration is completely unclear. In protostellar objects, it is
  believed that this reflects the different phases of formation of
  water and organics, with water being formed very early when the gas
  is not very cold, while the organics were synthesized later from frozen
  CO. It is possible that a similar process is at the origin of the different
  deuteration in meteorites and comets.  Alternatively,
  they may conserve a ``core'' of the
  protostellar molecular deuteration.
\item Molecular deuteration in a given molecule has a strong
  temperature dependence at low temperatures. This is true for both
  gas phase and icy grain surface formation.  In ice mantles the
  higher binding energy of deuteration atoms at $<$ 20 K leads to
  significant enhancements of deuterated species that form via
  hydrogenation reactions (see also Figure~\ref{fig:steve_deut}).
  These low temperatures correspond to what modelers predict for the
  cold midplane in the giant planet formation zone of the
  protoplanetary disk.  Measurements of the molecular deuteration in
  comets provide a potential tracer of the thermal conditions under
  which they formed.   For further discussion see
   Bockel{\'e}e-Morvan et al. (this volume)
\item Molecular deuteration in a given molecule has a strong
  temperature dependence at low temperatures. This is true for both
  gas phase and icy grain surface formation. In ice mantles the higher
  binding energy of deuteration atoms at $<$ 20 K leads to significant
  enhancements of deuterated species that form via hydrogenation
  reactions (see also Figure~\ref{fig:steve_deut}).
  Measurements of the molecular deuteration in comets provide a
  potential tracer of the thermal conditions under which they formed.
  For further discussion see Bockel{\'e}e-Morvan et al. (this volume).
\end{itemize}

\subsubsection{\label{sec:carmen}A model of the deuteration of water in the early stages of the evolution of the
  solar nebula}

\begin{figure}[!htb]
\center
\includegraphics[width=0.8\linewidth]{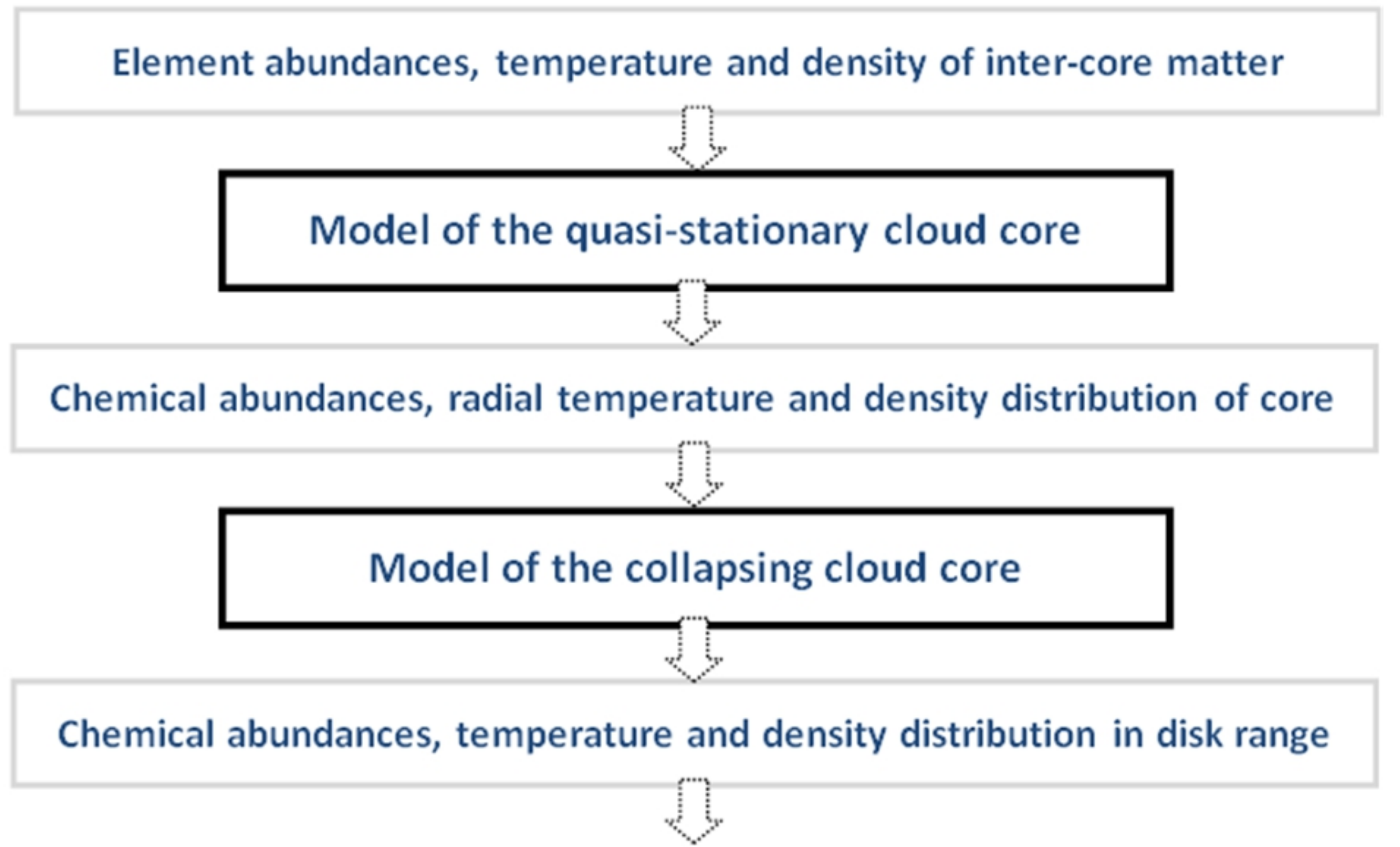}
\caption{\label{fig:tornow1} Flowchart
illustrating the evolution of the two precursor stages (the
quasi-stationary cloud core and the collapsing core) to the solar
nebula.  Evolution in these stages will affect the composition of
material incorporated into the solar nebula and hence the composition
of planetary bodies that form within it.}
\end{figure}

Molecular deuteration potentially provides a means of tracing the
history of solar system material and its links to the parent
molecular cloud. In this section we present the results of a specific model of
the evolution of the deuteration of water during the collapse of the parent
molecular cloud, since the composition of the protosolar nebula
depends in part on the chemical history of this core (Figure~\ref{fig:tornow1}).
The evolution of the core has been modeled recently by
\cite{aikawa12,wakelam14,taquet14} and \cite{tornow14a,tornow14b}. 

\begin{figure}[!htb]
\includegraphics[width=\linewidth]{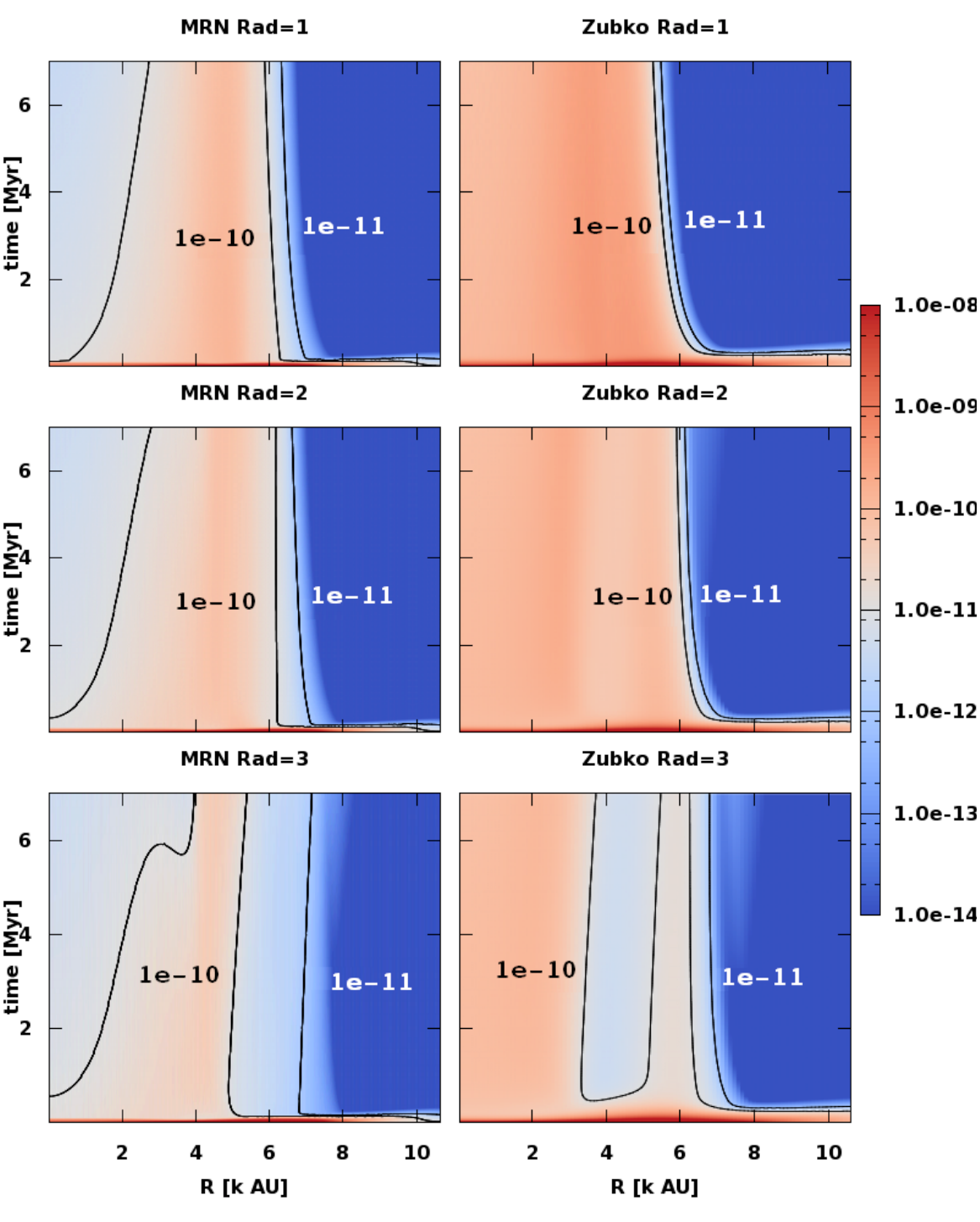}
\caption{\label{fig:Tornow_H2O}Relative gas phase abundance of ater
  versus time and radius (1 kAU = 10$^3$ AU).  The intensity of the
  standard interstellar radiation field (IRSF) is scaled by Rad = 1,
  2, 3.  Two grain distributions are used.  MRN refers to \cite{mrn}
  and Zubko refers to \cite{zubko}.  The mean grain size is 5.5
  $\times$ 10$^{-5}$ cm and 3.5 $\times$ 10$^{-5}$ cm for MRN and Zubko
  respectively.  The cosmic ray ionization rate is taken to be
  10$^{-17}$ s$^{-1}$.  The related
gas and grain chemistry is modeled using a combined code \citep{tornow14a}.}
\end{figure}

\begin{figure}[!htb]
\includegraphics[width=\linewidth]{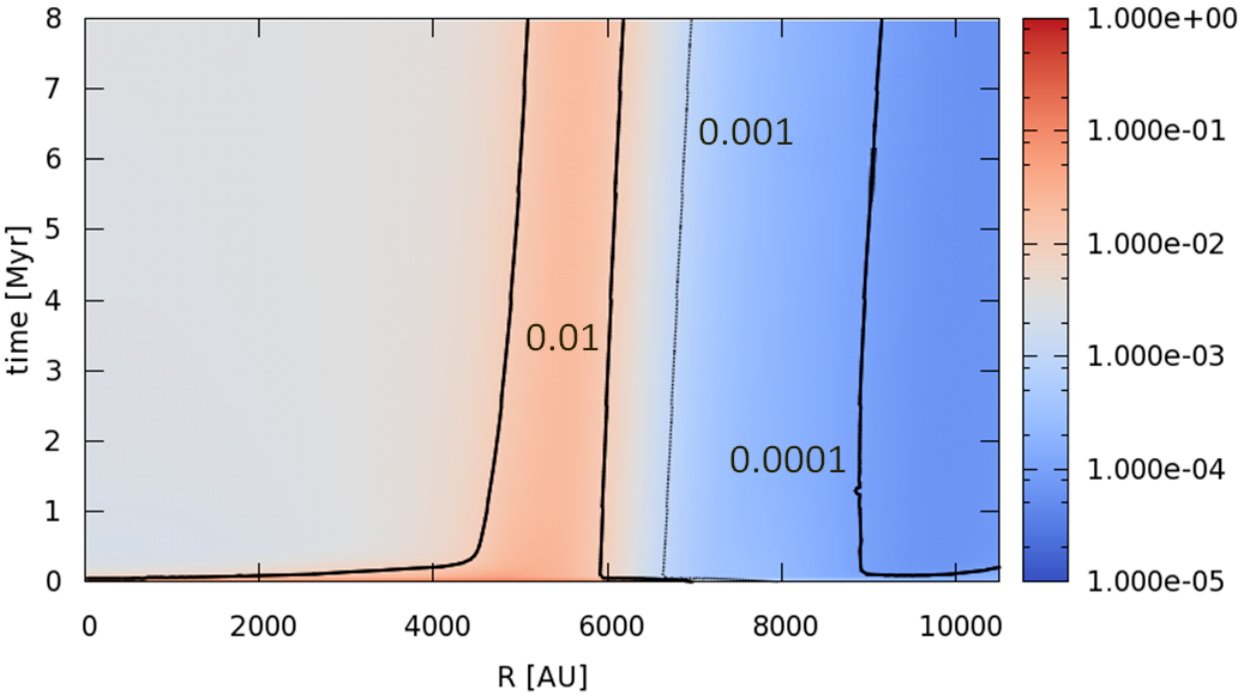}
\caption{\label{fig:Tornow_HDO}Temporal and radial HDO/H$_2$O
  distribution in the gas phase \citep{tornow14b} for the standard ISRF and MRN grain
  size (see Figure~\ref{fig:Tornow_H2O}).}
\end{figure}

In these models the cloud core is initially described
as a Bonner-Ebert sphere \citep{aikawa08} which is embedded in a
constant density gas-shell wiht a visual extinction of up to 4
magnitudes \citep{wakelam14}, and irradiated by an external UV field.
It is allowed to evolve for between 1 and 6 Myrs \citep{tornow14a}.
Both the gas phase\footnote{Nahoon{\_}public{\_}apr11.tar.gz at
  \url{http://kida.obs.u-bordeaux1.fr/models/}}  and 
grain\footnote{disk{\_}chemistry{\_}OSU08ggs{\_}UV.zip at
  \url{http://www.mpia.de/homes/semenov}} chemistry is followed.  Since the core is
only partially shielded, photo-chemistry has a clear effect on the
radial profile of the relative water abundance in the outer core
region (Figure~\ref{fig:Tornow_H2O}).  At the end of the
quasi-stationary core stage the temperature of the gas and dust
decreases to 8 -- 9 K in the center for boundary values between 13 --
15 K.  Consequently, water ice builds up on the cold grains, reaching
to a fairly constant relative abundance of $\sim$ 10$^{-4}$ within
10$^{5}$ -- 10$^{6}$ years.  In contrast, Figure~\ref{fig:Tornow_H2O}
shows a clear radial gradient for the gas phase water abundance.
Relatively low ratios (about 0.01 -- 10 ppm) of gas to dust phase
abundance were obtained for the outer core range with up to 25 ppm in
the inner region.  \cite{caselli12} determined higher ratios due to a
higher efficiency of water photodesorption; a consequence of the
larger grain size and smaller photodesorption yield used in the Tornow
et al. models.  If the deuterium
chemistry is included according to the method of \cite{albertsson13}
the temporal and radial distribution of the gas phase HDO/H$_2$O ratio
can be derived.  It reaches a maxiumum of about 0.02 in the medium
shielded core region, and a minimum of 10$^{-4}$ in the low shielded
outer boundary region (Figure~\ref{fig:Tornow_HDO}).  In contrast, the
dust phase ratio of HDO/H$_2$O is clearly larger and only varies
between 0.01 and 0.1.  Its maximum is located in the well shielded
inner core region at very early times ($\sim$ (2 --3) $\times$ 10$^{5}$
years).  

\begin{figure}[!htb]
\includegraphics[width=\linewidth]{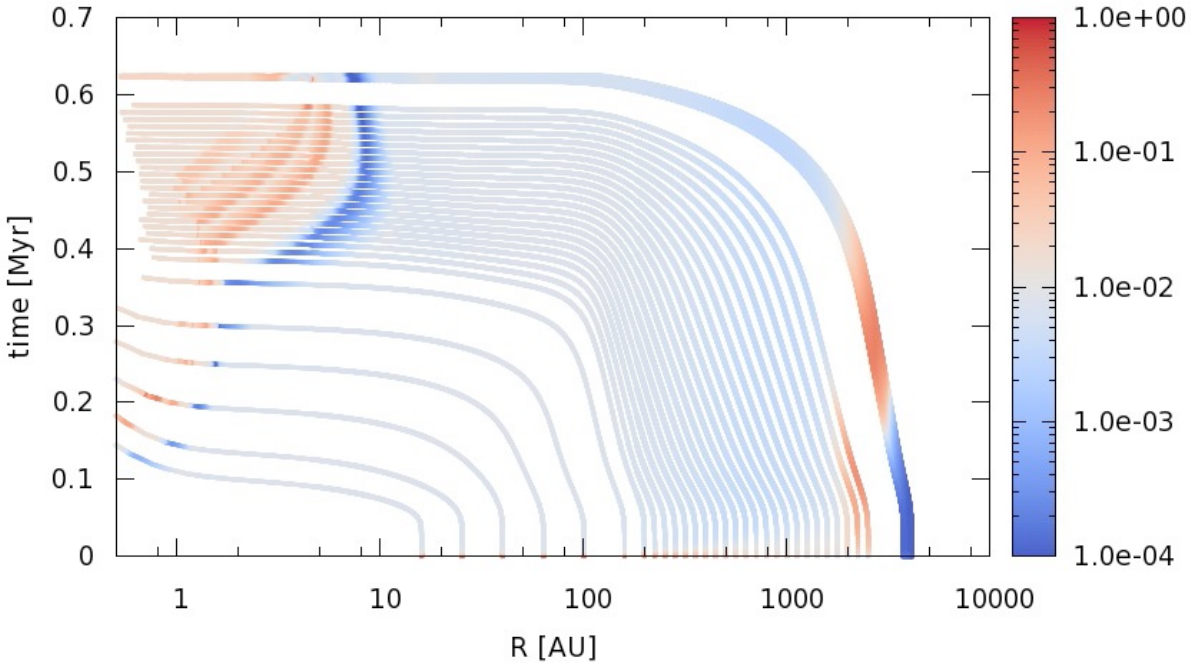}
\caption{\label{fig:Tornow_lagrange}HDO/H$_2$O ratio in the gas phase
  on Lagrangian paths.  At 1 AU the HDO/H$_2$O ratio varies between
  10$^{-3}$ and 0.1, while in the dust phase it is much smoother, and
  only varies between 0.03 and 0.05 \citep{tornow14a,tornow14b}.}.
\end{figure}

The collapse of the cloud core forms a protostar surrounded by an
optically thick disk and an outer envelope \citep{saigo08,st11}.
During this stage the chemical evolution can be followed along a
Lagrangian path from the outer core boundary to the center of the core
\citep{tornow14b}.  Figure~\ref{fig:Tornow_lagrange} shows the
HDO/H$_2$O ratio along each member of the Lagrangian path set.  Water
formed in the quasi-stationary core stage is transported into the
region of terrestrial planet formation.  Signatures of large-scale,
supersonic water inflows have been seen with Herschel \citep{mottram}
who determined that these flows occur on the core scale ($\leq$ 10$^4$ AU),
e.g., IRAS 15398 and L1527 or at least on the envelope scale ($\geq$ 3
$\times$ 10$^{4}$ AU), e.g., NGC1333-IRAS4A.
Based on their data Mottram et al. suggest an outside-in collapse, in agreement
with the model shown in Figure~\ref{fig:Tornow_lagrange}.

The gas phase deuteration of water varies widely, depending on
location and evolutionary time.  At 1 AU it ranges from 0.001 for 0.1
-- 0.2 Myrs, to 0.1 for 0.3 Myrs and to 0.03 for times later than 0.4
Myrs.  A minimum HDO/H$_2$O ratio near 0.001 extends from the radial
region around 0.1 AU at early collapse times to nearly 9 AU at later
times.  The predicted HDO/H$_2$O ratios of \cite{wakelam14} and
\cite{tornow14b} are slightly lower than those of \cite{aikawa12} but
larger than the ratios observed by \cite{persson}.  An
over-estimation of the water deuteration may be caused by a low
collapse temperature which would suggest perfect cooling.  Including
the ortho-to-para ratio of H$_2$ can also lead to lower HDO/H$_2$O
ratios \citep{taquet13,taquet14}.  In addition, water formed during
the quasi-stationary core stage must be destroyed by the UV radiation
of the interstellar radiation field, nearby stellar
sources or X-rays (producing secondary UV photons in a similar way as
cosmic rays) and UV photons from the protostar
\citep{france}.  As a result, photodissociation of deuterium-rich
water molecules (HDO + h$\nu$ $\longrightarrow$ D + OH, or H + OD),
together with 1-D radial or vertical mixing enable a subsequent
reformation of water by hot neutral reactions in the inner disk plane
\citep{furuya13}.

\subsection{\label{sec:c_frac}Fractionation of carbon and nitrogen in the solar nebula}
  Isotopic enrichment of $^{15}$N in cometary organics (the nitriles
  CN and HCN) and ammonia is commonly observed
  \citep[][Bockel{\'e}e-Morvan et al., this issues]{jehin09}.
Chemical models have shown that fractionation in 
 low-temperature ion-molecule reactions can produce large $^{15}$N
  enrichments in interstellar clouds \citep{rc08,wirstrom12}.
However, a recent study by \cite{roueff15} has demonstrated that
several key reactions may in fact possess energy barriers, leading to
doubts as the efficacy of ion-molecule processes in N fractionation.
Of particular interest for comets is that
  $^{15}$N fractionation in the outer nebula should be similarly
  impaired by low values of the cosmic ray ionization rate, $\zeta$. 
Thus, as found by \cite{cleeves14a}
   for water deuteration, the low $^{14}$N/$^{15}$N ratios commonly
  measured in both Oort Cloud and Jupiter Family comets is strongly
  suggestive of an interstellar origin.

  The $^{12}$C/$^{13}$C ratios measured thus far in simple cometary
  molecules (CO, HCN, CN) are solar \citep{jehin09} and
  Bockelee-Morvan (these proceedings).  In regions of the nebula where
  water has condensed, the gaseous C/O elemental ratio can increase to
  order unity \citep{oberg11}. Because water, OH and O$_2$ recycle
  carbon nuclei back into CO \citep{langer84}, ion-molecule
  fractionation in regions with C/O$\sim 1$ will act instead to
  channel carbon nuclei into simple organic molecules.  Thus, as the
  sole carbon fractionation process involves the $^{13}$C$^+$ exchange
  reaction with $^{12}$CO \citep{langer84}, CO condensation may in
  fact be the reason for the almost uniformity of cometary
  $^{12}$C/$^{13}$C ratios measured to date.  On the other hand, the
  condensed CO is predicted to produce distinctive $^{12}$C/$^{13}$C
  fractionation patterns in the organic molecules formed in the
  grain-surface reaction scheme of Figure~\ref{fig:stern}
  \citep{charnley04}.  These involve molecules whose $^{13}$C
  isotopologues have not yet been detected in comets (e.g., H$_{2}$CO,
  CH$_{3}$OH) but for which the predicted fractionation has been
  confirmed by observations of protostellar envelopes
  \citep{wirstrom11}.  In this case, once again, ion-molecule pathways
  to $^{13}$C fractionation in the nebula are predicated on the
  presence of significant flux of cosmic rays.
  
If ion-molecule reactions indeed play a greatly reduced role in
the cooler parts of the disk then the major sources of nebular
fractionation will involve UV photochemistry.  Isotope selective
photodissociation of CO has been proposed as an explanation of the
oxygen isotope ratios measured in meteorites \citep{ly}.
Isotopic enrichment of $^{13}$C, $^{15}$N and D though
isotope-selective photodissociation of isotopologues of CO, N$_2$
and HD has also been considered \citep{woods_willacy09,heays14,cleeves14a}.
However, these processes will
only occur in thin upper layers of the disk and so will be
negligible at the midplane.  Thus, to be viable, these processes
requre efficient downward vertical mixing in the disk.  On the other
hand, ISP occurring in the over layers of the presolar core, where the
interstellar UV field is attentuated, could in principle produce
isotopic anomalies in the oxygen and carbon isotopes
\citep{yk04,lee08,cr09}.  This environment could also ultimately
produce significant fractionation in nitrogen-bearing species but this
has not yet been demonstrated quantitatively. 

\subsection{\label{sec:ohem}Solar nebula chemistry and the origin of oxygen
  isotopic anomalies in the solar system}

\begin{figure}
\includegraphics[width=\linewidth]{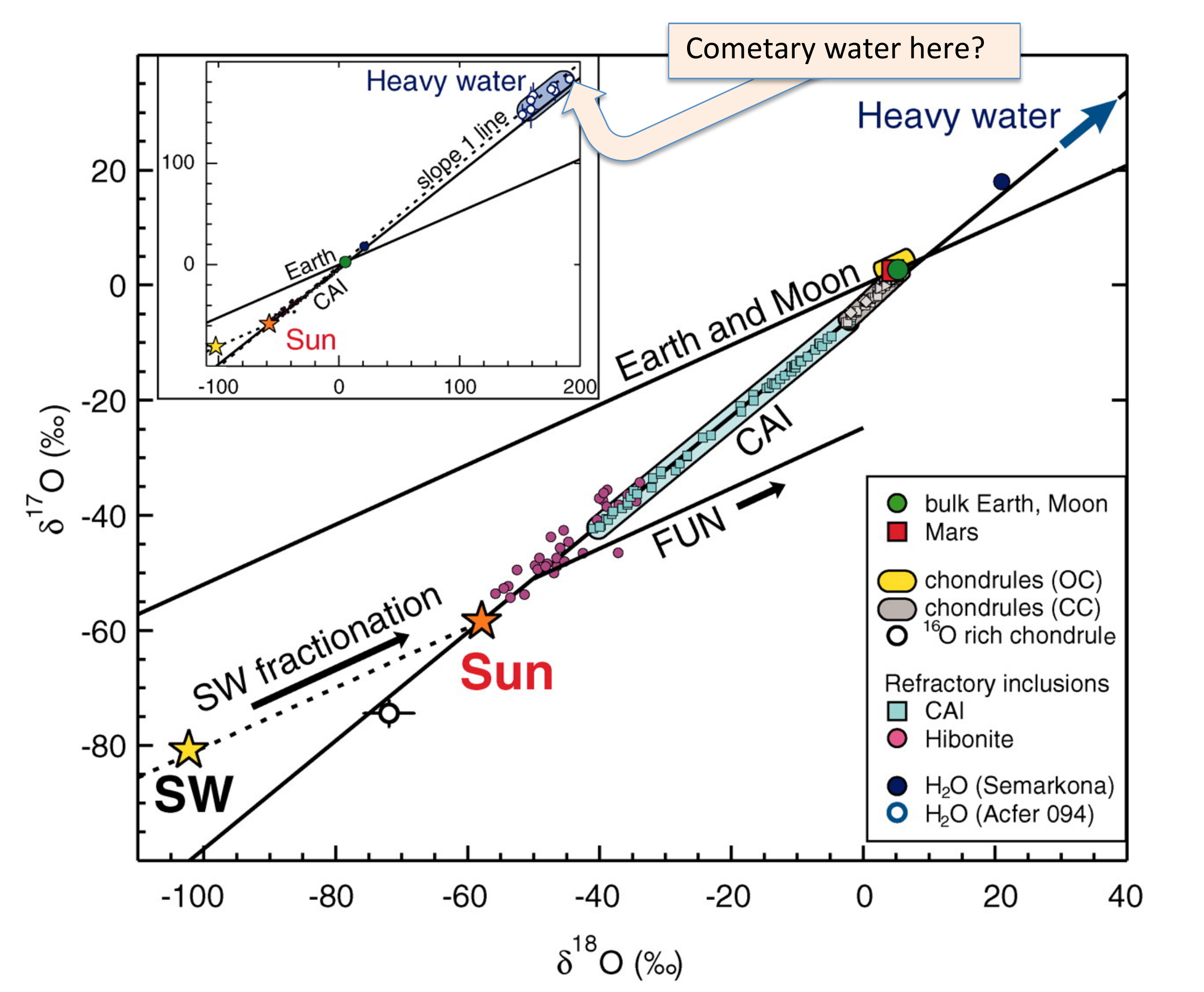}
\caption{\label{fig:claudia2}A three-isotope plot for oxygen in the
  solar system.  The Earth and Moon lie along a fractionation line
  with slope 0.52.  This line is consistent with fractionation of a
  single homogenous source, by processes that depend on the mass of
  the isotopes, so that $^{18}$O is nearly twice as fractionated as
  $^{17}$O.  Data from meteorites and the Genesis mission falls along
  a slope-1 line indicating they contain material that has undergone
  some kind of mass-independent fractionation.  The Genesis results
  suggest that much of the rocky material in the inner solar system
  was enriched in $^{17}$O and $^{18}$O relative to $^{16}$O before
  the accretion of the first planetesimals.  Hence rocky planets are
  not made of the average nebular material that formed the Sun.  
From \cite{mckeegan11}  (reprinted with permission).}
\end{figure}

Oxygen is a very common element that is a major constituent of many
minerals and rocks.  It exists in three stable isotopes: $^{16}$O,
$^{17}$O and $^{18}$O.  The fractionation between these isotopes
is not consistent across all solar system samples, as might be
expected if the oxygen isotopes were homogeneously distributed across
the solar nebula, and if only mass-dependent fractionation processes
were active.  Figure~\ref{fig:claudia2} is a three-isotope plot of
the oxygen isotopic compositions of primitive solar system materials
such as chondrites, chondrules, shows that their relative isotopic
compositions are different compared to those of terrestrial samples \citep{mckeegan11,yurimoto07,clayton93}
The terrestrial line is characteristic of
mass-dependent fractionation, where $^{18}$O is twice as fractionated
as $^{17}$O relative to $^{16}$O, whereas the steeper, slope -1 line, of the
primitive materials suggests it is the result of mass-independent
fractionation.  

There are three possible mechanisms that could achieve this
mass-independent effect \citep{yurimoto07}.  Firstly the
isotope anomalies could represent an inherited heterogeneity in solar
system materials, where the solids are rich in $^{16}$O and the gas in
the rarer isotopes.  This could occur either from nucleosynthesis in
other stars \citep{clayton73} or from nuclear reactions triggered by
energetic particles produced by the Sun or by galactic cosmic rays
\citep{lee78}.  These nuclear processes could produce fluctuations
in the $^{16}$O abundance.  The lack of observed isotopic anomalies in other
elements suggests this mechanism is unlikely \citep{clayton93}.
Additionally, $^{16}$O presolar grains are rare in meteorites
\citep{nagashima04,nittler03}.

A second suggestion is chemical fractionation within the solar
nebula.  Differences in the reaction rates of symmetric molecules,
e.g.\ $^{16}$O$^{16}$O, compared to asymmetric ones, e.g.\
$^{16}$O$^{17}$O, can lead to the observed isotopic distributions,
since the most abundant isotope is almost entirely contained in the
symmetric molecules, while the rarer isotopes are in the asymmetric
ones \citep{ht83}.  However, the exact reactions by which this effect
might be achieved are uncertain \citep{yurimoto07}

The final possibility is isotope-selective photodissociation of CO, either in the parent cloud
\citep{yk04} or at the surface of the protosolar disk \citep{young,ly}
could also result in enhancements of the rare isotopes of oxygen.  
CO dissociates when irradiated by far-ultraviolet photons with
wavelengths between 91.2 nm and 110 nm.  The dissociation process
occurs in two steps, with CO first entering a bound excited state
before dissociating.  As a result the CO absorption spectrum consists
of a large number of narrow lines at specific wavelengths determined
by the vibrational and rotational levels involved.  The change in mass
between C$^{16}$O and its isotopologues C$^{17}$O and C$^{18}$O
slightly shifts the lines in the absorption spectra, and consequently
the spectra for the different isotopologues do not overlap.  Hence
when CO is dissociated, either in the protostellar disk or in the
molecular cloud core, the different isotopologues do not shield each
other \citep[see, for example,][]{visser}.  
C$^{16}$O will become optically thick closest to the surface,
in a region where C$^{17}$O and C$^{18}$O can still dissociate.  
When coupled with turbulent mixing that can transport the $^{17}$O and
$^{18}$ towards the colder, shielded midplane of the disk this can
result in the formation of water molecules that are enhanced in the
rarer isotopes and which could later be incorporated into CAIs and
other meteoritic material.

The in situ measurements that Rosetta will make will directly probe
the isotopic composition of the volatile component of a comet.  This
will test the viability of the potential causes of the oxygen isotopic
anomalies and provide new clues as to the origin of the cometary
material and the conditions under which it formed.

\section{\label{sec:n}The origin of the nitrogen deficiency in comets}

Nitrogen is one of the most abundant elements in the PSN
\citep{2009ARA&A..47..481A}. Its abundance was measured in the giant
planets atmospheres in which its stable thermodynamic form is
NH$_{3}$. The nitrogen abundance is found moderately to substantially
enriched, compared to its solar value, in Saturn and Jupiter
\citep{2014arXiv1404.4811M}, but is loosely constrained in Uranus and
Neptune \citep{dp1,dp2,mad}. A puzzling revelation was the apparent
severe depletion of nitrogen in comets. Mass spectrometry measurements
of 1P/Halley's dust composition allowed to find that the
nitrogen-to-oxygen ratio is depleted by a factor $\sim$3 with respect
to the solar value \citep{jk91}. Subsequent analysis showed that the
ice part of the comet is depleted in nitrogen by a factor up to 75
\citep{w91}. This apparent depletion appears controversial because the
N$_2$ molecule, thought to be the main nitrogen-bearing volatile in
the whole PSN \citep{lewis,mou02}, has a molecular mass close to that
of the very abundant CO molecule, implying that the mass spectrometer
aboard the Giotto spacecraft did not have enough resolution to
separate the two molecules \citep{e87}. 

More recently, \cite{c2000} used the 2DCoude spectrograph on the 2.7-m
Harlan J. Smith telescope of McDonald Observatory to observe the
N$_2^+$/CO$_2^+$ in comets 22P/1995 S1 (deVico) and C/1995 O1
(Hale-Bopp), and found values around $\sim 1\times 10^{-4}$. These
values hint at a severe depletion in N$_2$ since the N$_2^+$/CO$^+$
ratio is usually equal to N$_2$/CO. This is surprising specially that
N$_2$ and CO have very close condensation temperatures
\citep{fs}. Moreover, efforts were made to find N$_2$ using the Far
Ultraviolet Spectroscopic Explorer, but with no success
\citep{bm04}. All these observations, and many more, motivated
theorists to try to explain this depletion. 

In \cite{iro}, the authors assumed that volatiles are accreted by
comets as clathrates \citep{lunine}, and explain the nitrogen
depletion by the low efficiency of N$_2$ trapping in water ices. One
of this work's conclusions is that N$_2$ can be trapped only if water
ice is present in abundance higher than 2.8 $\times$ solar value. This
is due to the relative ease by which CO can be trapped in comparison
with N$_2$, so this last can be clathrated only if there is enough
water left after all the present CO was captured. Since N$_2$ is
trapped in crystalline water ice at temperatures lower than CO
\citep{m2002}, their work implied that comets might have formed at
temperatures higher than 45 K, allowing the clathration of CO but not
N$_2$. This temperature lower limit is one of shortcomings of this
model since in many protoplanetary disks models, temperature is
thought to decrease below this value \citep{hg}. The recent
observation of the CO iceline in TW Hya also supports these models
\citep{qi}. Another caveat is the reliability of the \cite{iro}'s
determination of the N$_2$/CO ratio in clathrates since the potential
parameters describing the guest--clathrate interaction in their
statistical model correspond to guest--guest interactions (case of
pure solutions) and not to the guest--water interactions usually used
in such models \citep{m2012}. Also, the model is unable to explain the
thick N$_2$ ice cover on Triton and Pluto, both bodies thought to be
formed in the comets region \citep{lellouch}. 

Motivated by the shortcomings of \cite{iro}, \cite{m2012} introduced
another statistical model based on an appropriate set of potential
parameters and that seems to address these issues. In this model,
volatiles are also accreted as clathrate-hydrates, except N$_2$ which,
due to its poor trapping propensity, is assumed to condense at
$\sim$20 K in the PSN and subsequently trapped in pure cristalline
form in comets. Later radiogenic heating by the decay of short-lived
nuclides will heat up these bodies to temperatures enough to liberate
the trapped N$_2$, but not CO. In Triton and Pluto, the gravity is
sufficiently important to prevent the gaseous N$_2$ from escaping, and
let it condense on the surface and form the observed cover. In the
case of comets however, the gravity is negligible and the gaseous
N$_2$ will escape, leaving comets depleted in nitrogen. A shortcoming
of this model is that no radiogenic particles were found by stardust
\citep{2008EM&P..102..447F}. 

A recent model, applied to the description of Uranus and Neptune
formations, also provides an alternative solution. \cite{mad} proposed
that the diffusive redistribution of vapor will remove the gaseous
N$_2$ from the region below its iceline, possibly explaining the
probable nitrogen depletion in Uranus and \& Neptune if they formed in
this region. One can invoke a similar argument for the nitrogen
depletion in comets. In this case, the CO--rich comets will need to
form between the CO and N$_2$ icelines, in order to accrete CO in
solid form (since its vapor will also be depleted below its iceline)
but not N$_2$. The relatively small distance between the two icelines
($\sim$4--5 AU) is however a caveat for this model since it imposes
the formation of all comets in this restricted area of the outer PSN.

Finally, all these models took the N$_2$ depletion at face value. As
discussed above, the N$_2$ molecule's mass is very close to CO, and
the two molecules can be distinguished only using an instrument with a
very high spectral resolution, such as ROSINA on board the Rosetta
spacecraft \citep{rosetta}. If this (or another) instrument is able to
detect important amounts of N$_2$ in a comet, this might resolve at
least part of the problem. Since important chemical differences are
known to exist between Oort Cloud comets and Jupiter Family comets
\citep{mumma}, positive detections or definitive exclusion of N$_2$ in
both types of populations are needed before drawing final conclusions.

\section{\label{sec:spin}Spin temperature as a cosmogonic indicator in comets}

The volatile composition, D/H ratio in molecular volatiles and the
spin temperature, as given by the ortho-to-para ratio in water or
other molecules, are considered to be possible cosmogonic indicators
for comets and as such may be interpreted as providing information
about the formation history of the molecules in question. In this
section, we discuss spin temperature as a possible cosmogonic
indicator for comets.

Molecules that contain identical hydrogen nuclei (like H$_2$O and
CH$_4$) display nuclear spin isomers. Water, for example, has two spin
species referred to as ortho-H$_2$O if the nuclear spins of the
hydrogen atoms are parallel and para-H$_2$ if they are
anti-parallel. The lowest energy level of para-H$_2$O lies 23.8
cm$^{-1}$ ($\sim$34~K) below the lowest ortho level, so when water is
formed at temperatures below about 50~K, the formation of para states
is preferred and the ortho-para ratio (OPR) is less than the statistical equilibrium
value of 3. Likewise for methane, statistical equilibrium corresponds
to abundance ratios A:E:F of 5:2:9 and statistical equilibrium is
reached for temperatures above about 60 K.

\begin{figure}
\center
\includegraphics[width=0.8\textwidth]{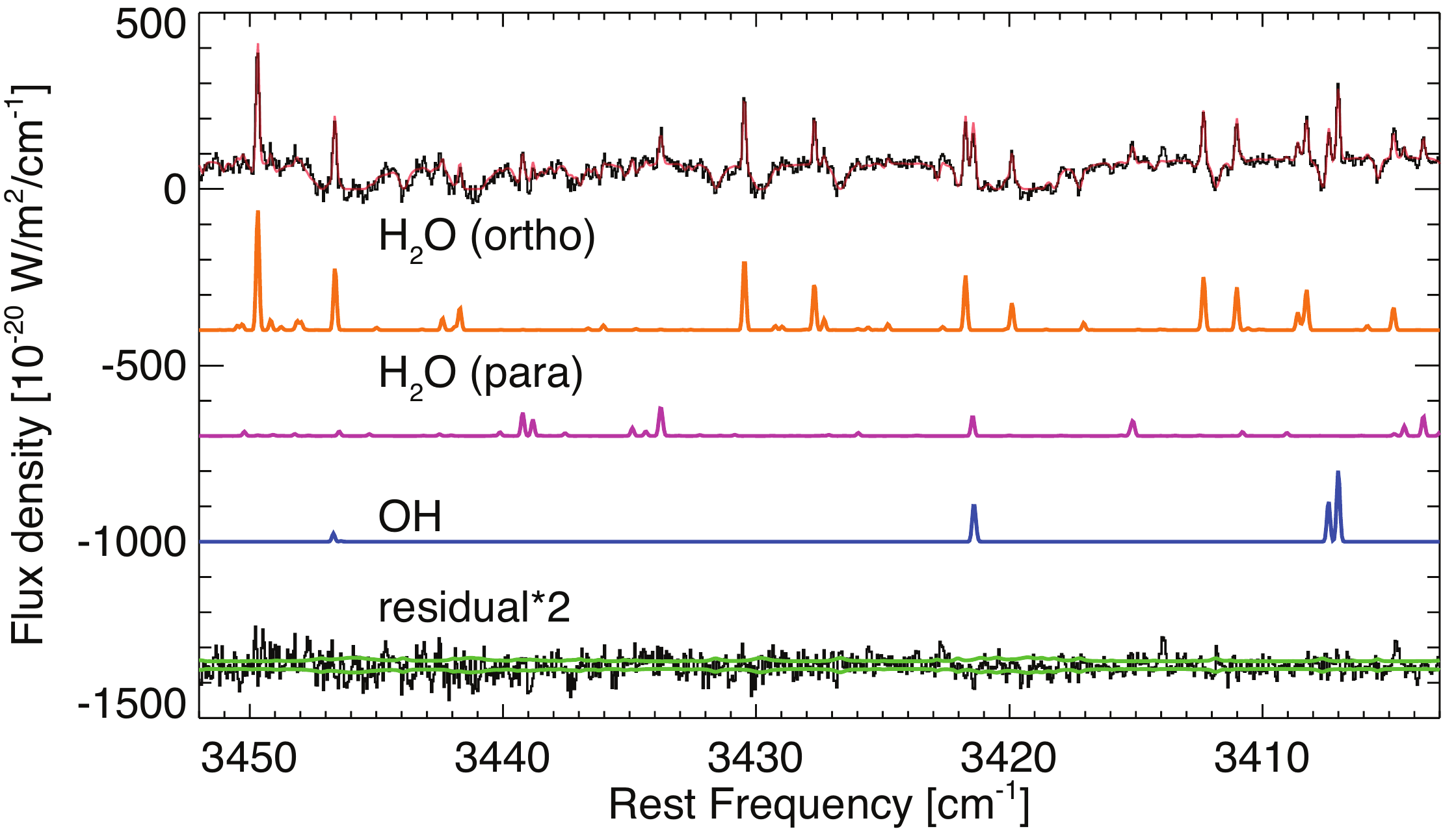}
\caption{Spectrum of comet C/2007 N3 (Lulin) on 31 Jan 2009 with
  the sum of the telluric and best-fit synthetic fluorescent emission
  model (red) overplotted. The best fit water ortho (orange) and para
  (purple) models are shown. Also shown is OH prompt emission
  (olive). To illustrate the quality of the fit, the residual (scaled
  by a factor of two) with the best-fit models subtracted is
  shown. The 1-sigma error envelope is overplotted in green. The best
  fit is for statistical equilibrium (T$_{spin}$ $>$ 60~K).}
\label{fig:lulin_h2o}
\end{figure}

\begin{figure}
\center
  \includegraphics[width=0.8\textwidth]{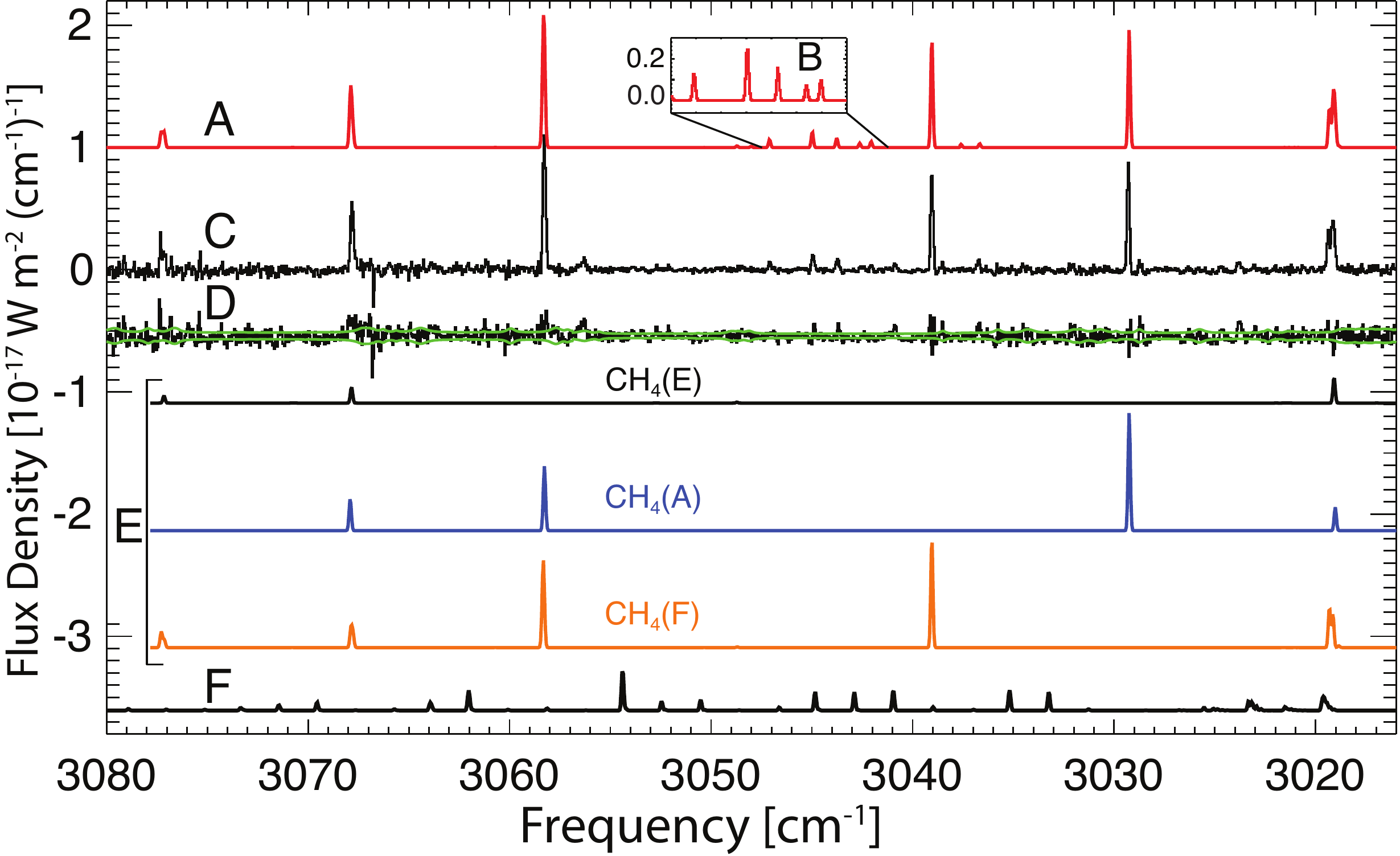}
  \caption{Residual spectrum of comet C/2007 N3 (Lulin) on 31 Jan 2009
    (C) with the sum of the best-fit synthetic fluorescent emission
    model (red) overplotted (A) and offset vertically for clarity. The
    OH prompt emission is shown in the inset (B). To illustrate the
    quality of the fit, the residual (scaled by a factor of two) with
    the best-fit models subtracted is shown in (D) with the 1-sigma
    error envelope overplotted in green. Beneath the residuals,
    individual fluorescent emission models are shown for the three
    spin components of CH$_4$ (E) and CH$_3$D (F) expanded vertically
    for clarity. The best fit is for statistical equilibrium
    (T$_{spin}$ $>$60~K). Adapted from \cite{gibb12}.}
\label{fig:lulin_ch4} 
\end{figure}

\begin{figure}
\center
\includegraphics[width=0.8\textwidth]{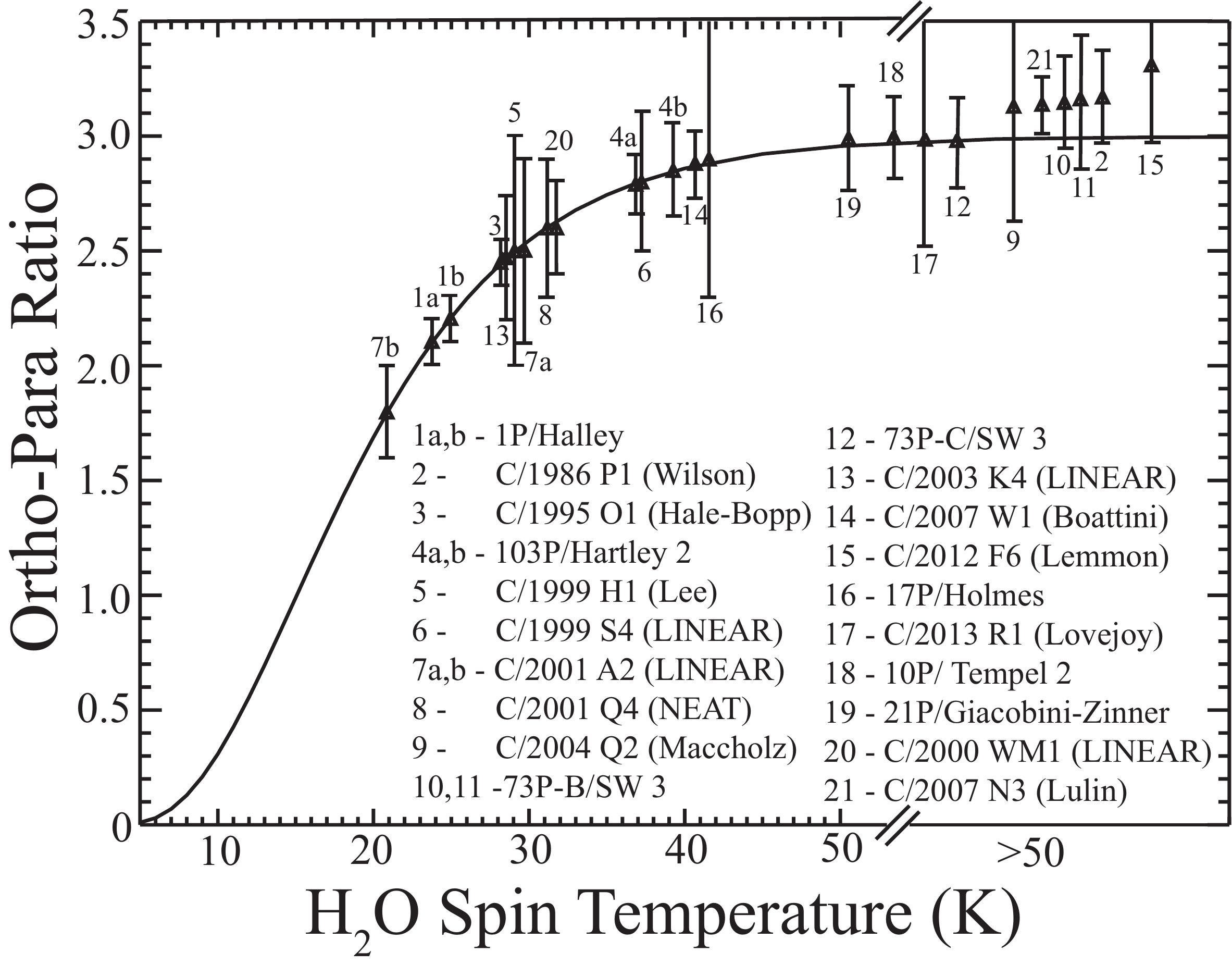}
\caption{OPR vs Spin Temperature in water with measurements for
  individual comets. Individual OPR measurements are from Mumma et
  al. (1988:1a, 1b, 2); Crovisier et al. (1997: 3, 1999: 4a, b); Dello
  Russo et al. (2005: 5, 6, 7a, 7b, 2007: 11, 12, 2008: 16);
  Kawakita et al. (2006: 8); Bonev et al. (2009: 9, 2008: 10);
  Woodward et al. (2007: 13); Villanueva et al. (2011: 14); Paganini
  et al. (2014: 15, 2015: 17, 2012: 18); DiSanti et al. (2013: 19);
  Radeva et al. (2010: 20); Gibb et al. (2012: 21).}
\label{fig:water_opr}
\end{figure}

OPRs have been routinely measured in water in comets since the
apparition of comet 1P/Halley \citep{mumma87}. This is currently
done by fitting synthetic fluorescent emission models independently
for each spin state as shown in Figures~\ref{fig:lulin_h2o} and
\ref{fig:lulin_ch4}. Corresponding spin temperatures are determined by
placing the OPR on a theoretical curve such as that for water shown in
Figure~\ref{fig:water_opr}. OPR is also commonly inferred for ammonia
by measuring the OPR for NH$_2$, thought to be formed primarily via
photodissociation of ammonia \citep{kawakita04,shinnaka11}. 
More recently, spin temperature has been measured in
cometary methane for comets C/2007 N3 Lulin \citep{gibb12}, C/2007
W1 \citep{villanueva11}, C/2004 Q2 \citep{bonev09,kk09}, 
and C/2001 Q4 \citep{kawakita05}.

The meaning of spin temperatures has been the subject of much
debate. For an isolated molecule of water or methane in the gas phase,
for example, nuclear spin conversion is strongly forbidden. For this
reason, spin temperature has been considered a cosmogonic indicator in
comets that can provide information on the formation temperature
\citep{crov84,mumma87}. Observational evidence supports
the suggestion that OPRs do not change once volatiles are released
into the gas phase. For example, recent results mapping the spin and
rotational temperatures in cometary comae are consistent with a
scenario in which the OPR is unchanged by coma processes \citep{bonev08}.
Also, the OPR for NH$_2$ does not correlate with the
heliocentric distance (and hence rotational temperature) of the comet,
nor does it correlate significantly with abundances of other volatiles
\citep{shinnaka11}.

However, it has also been suggested that the spin temperature of water
depends on the evaporation history of the ice \citep{bunt08}.
\cite{hama11} found experimentally that the OPR of
water desorbed from ice after being vapor deposited at 8~K or formed
via irradiation of a CH$_4$/O$_2$ mixture at 8~K was consistent with
statistical equilibrium, suggesting that OPR may equilibrate during
the thermal desorption process or that energetically produced water
does not undergo nuclear spin conversion (to a lower OPR) in water
ice. Both formation processes for water in the \cite{hama11}
experiment were energetic, and the authors note that the spin
temperature of water molecules produced by non-energetic processes
must also be studied before definitive conclusions can be drawn,
especially since cometary volatiles may have formed by low-energy
processes. Nonetheless, such studies raise the possibility that
cometary OPRs may not, in fact, be cosmogonic.

Can the measurements of OPR in comets shine any light on this
question? We show the measurements of cometary OPR in water to date in
Figure~\ref{fig:water_opr}. Water was chosen due to the greater number
of OPR measurements than for other molecules in comets. Early
measurements suggested a clustering near T$_{spin}$ $\sim$ 30~K,
leading many authors to conclude a formation temperature of $\sim$30 K
for a large population of comets. Similar conclusions have been drawn
from studies based on CH$_4$ \citep{kawakita06} and NH$_2$
\citep{shinnaka11}, particularly given the lack of correlation
between spin temperature and other properties of the observed comets
\citep{shinnaka11}. However, it must be noted that among the 20
comets measured to date, nearly all of them are within 1- or 2-sigma
of the statistical equilibrium value, excepting the measurements
reported for 1P/Halley, one of the two measurements for C/2001 A2, and
C/1995 O1 (Hale-Bopp). Likewise, half of all measurements are at or
above the statistical equilibrium value. It is also notable that no
measurements exist below a spin temperature of 20~K, and this
observation holds for methane and ammonia spin temperatures as well.

Possible interpretations of spin temperatures measured in comets are
therefore (1) that it is a cosmogonic indicator that gives the
formation temperature of the molecule in question and therefore
perhaps the formation conditions of the comet, (2) that it is a
measurement that, while unchanged by coma processes, is set by
processes occurring in the solid ice during or prior to
sublimation. Current data are not sufficient to draw firm conclusions
at this point. Even if further observations result in most comets
consistent with statistical equilibrium, we may still be sampling
cosmogonic material since it is known that comets contain material
processed at high temperatures, such as crystalline silicates, which
are interpreted to originate from turbulent radial mixing in the solar
nebula \citep{bm02}.  Mixing of volatile material
that formed at high temperatures in the inner disk to the comet
forming region would have resulted in a higher OPR than for
uncontaminated material that formed in the cold midplane. Clearly a
great deal more work needs to be done to understand the significance
of molecular spin states in cometary material.

%




\section{\label{sec:na}On the origin of neutral sodium in comet tails}

The presence of sodium D line emission has been confirmed in a large
number of comets close to perihelion since it was first reported more
than a century ago \citep{bredichin}. Observations of comet C/1995O1
Hale-Bopp during the spring of 1997 led to the discovery of a new tail
connected with the sodium D line emission. This neutral sodium gas
tail is entirely different from the previously known ion and dust
tails, and its associated source is unclear. Up to now, only physical
reasons have been advanced to rationalize its origin. It has been
proposed that this third type of tail is shaped by radiation pressure
due to resonance fluorescence of sodium atoms \citep{cremonese}.
Further possibilities were considered and then rejected, as
photo-sputtering and/or ion sputtering of nonvolatile dust grains \citep{Ip_Jorda}.
In the same way, collisions between the cometary dust
and very small grains were also questioned \citep{Ip_Jorda}. The
scenario presented in this review is completely different since
entirely based upon chemical grounds. It is shown that the Na$^+$
ions, washed out of the refractory material during the hyration phase,
loose their positive charge to evolve progressively into neutral
species with their migration towards the surface of cometary ices.  

\cite{ellinger} follow the chemical path of sodium, starting
from the incorporation into the ice until the final transformation
into a neutral atom when released from the sublimating crystalline
(see Section~\ref{sec:h2o}) cometary ice.  
The modeling of bulk,  surface structures and reactivity of ices 
has been essentially developed for environmental applications \citep{casassa02,casassa05,calatayud}
Levering on these  results, high-level numerical simulations based on
first principle periodic density functional theory (DFT) were used to
describe the solid structure of the ice in the form of apolar
hexagonal ice \citep{bussolin} composed of bi-layers of water
molecules.   

A code, specifically designed for the study of periodic systems,
namely, the Vienna {\it ab initio} simulation package (VASP)
\citep{kresse94a,kresse94b}
was used to carry out all the calculations.  
The generalized gradient approximation (GGA) was employed in the form
of the Perdew and Wang exchange-correlation functional \citep{perdew}
coupled to the Grimme correction \citep{grimme} to take care of
the long-range van der Waals interactions.  
The evolution of the ionic character of the sodium atom as it reaches
the ice surface was obtained  by a topological analysis \citep{ss94}
performed by means of the TOPMOD package \citep{llusar} and its
recent extension to periodic systems \citep{kp11}.

A first calculation on isolated  Na(H$_2$O) showed that the
interaction of Na with one H$_2$O is similar to that between two
H$_2$O molecules in the H$_2$O dimer ($\approx$ 0.3 eV), which gives a
hint  in view of a possible replacement of a molecule of water by a
sodium atom in organized structures.  
In liquid water, microsolvatation of sodium atoms has been studied on
small clusters of Na(H$_2$O)n (n=1-6) to mimic the water
environment. Both the structures in which Na is positioned on the
surface and the encapsulated structure in which Na is surrounded by a
solvation shell are minima on the potential surface. They are very
close in energy \citep{hashimoto}

\begin{figure}[!htb]
\centering 
\includegraphics[scale=0.4]{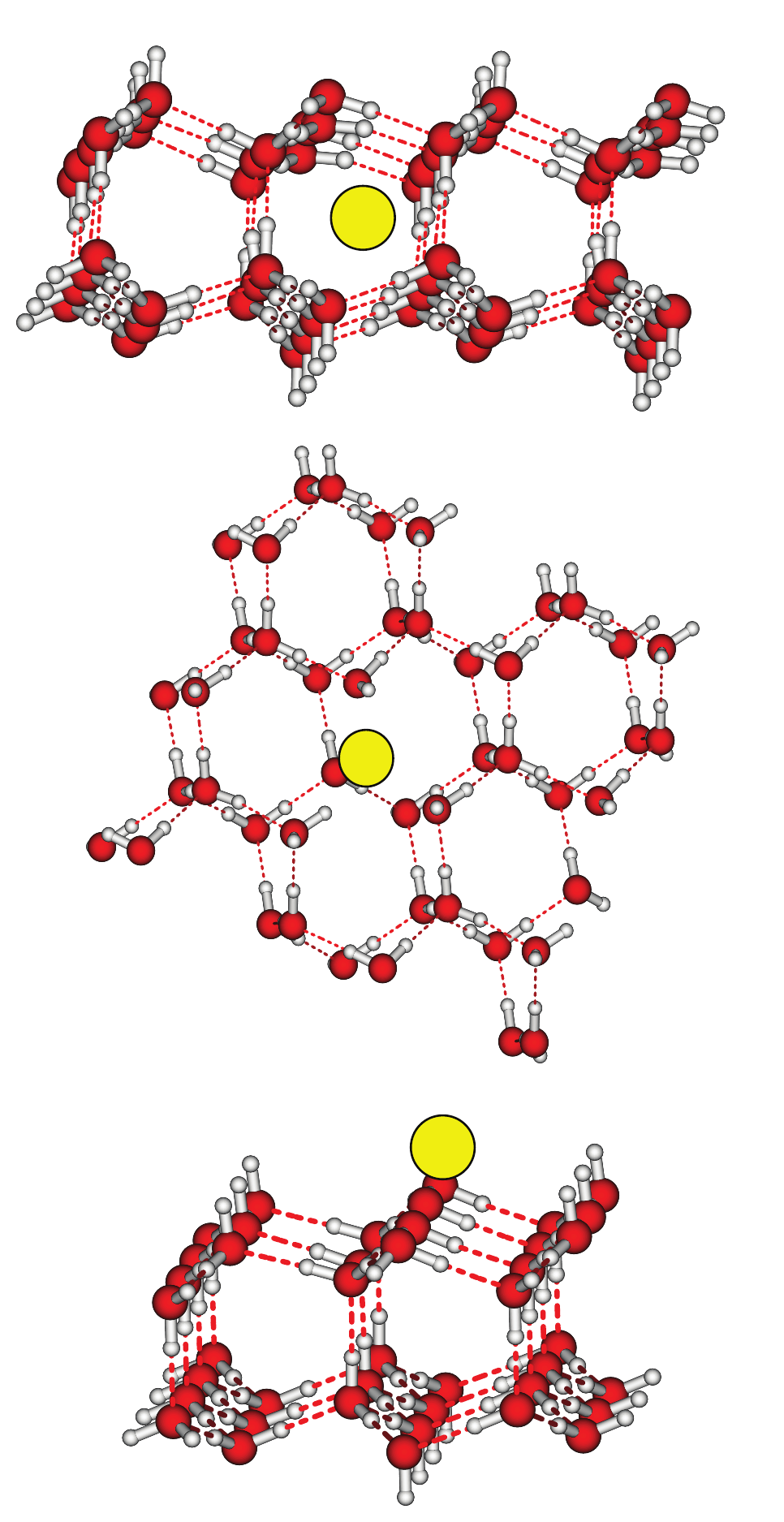}

\caption{\label{fig:na}Typical structures of Na-H$_2$O$_{ice}$ arrangements. Top: Na in the cavity inside hexagonal ice. Middle: Na replacing H$_2$O in the crystal. Bottom: Na adsorbed on the surface of the ice. {\it Color code} Yellow (Na); red (O) ; white (H).} 
\label{}
\end{figure}

In the ice, the periodic structure of the crystal imposes strong
geometrical constraints. The most stable structure, not found in the
liquid, is the one where Na takes the place of a water molecule in the
lattice (middle structure in Figure~\ref{fig:na}) when the ice is formed
($\Delta$E $\approx$ 1 eV).   
Contrary to  microsolvatation, the stability  of Na encapsulated in
the bulk of the crystal ($\Delta$E $\approx$ 0.5 eV),  shown in the
top structure in Figure~\ref{fig:na}) differs from that  of the adsorbed  Na
according to the position on the surface. For example, the adsorption
energies vary from ($\Delta$E $\approx$ 0.6 to $\approx$ 0.5 and
$\approx$ 0.1 eV) when the adsorption takes place over the center of a
surface hexagon, over a surface oxygen (bottom structure in Figure~\ref{fig:na})
and over an OH dangling bond, respectively.  

The most relevant point to this study is that the charge q$_{Na}$ on
the sodium atom depends strongly on the position with respect to the
ice surface, as does the adsorption energy.  When constrained to stay
in the cavity inside the hexagonal lattice, Na remains largely ionized
(q$_{Na}$ $\approx$ +0.9)  which is close to the value of $\approx$
+0.8  when Na takes the place of an H$_2$O molecule in the bulk. When
Na is adsorbed on top of an oxygen of the surface layer,
q$_{Na}$$\approx$ +0.2, a value similar to that found for the isolated
Na(H$_2$O) complex.  Then, as the ice is progressively eroded by the
radiation field, Na eventually reaches the surface where it can desorb
as a neutral atom when the upper layer of the ice vaporizes.  


\section{\label{sec:summary}Summary}

The observed compositions of comets are the end product of a chemical
history that began in the cold parent molecular cloud core.  The
change in physical conditions experienced by the core material as the
core collapsed and the protosun and protosolar disk formed and evolved
led to chemical changes that are reflected in the cometary volatiles
that we observe today.  Each stage of evolution left its imprint in the mixing ratios
of molecules observed in comets and disentangling their effects
requires interaction between comet and disk
researchers.   The combination of the two fields of study
provide a powerful tool in understanding the drivers that determine
comet composition and conversely what the variations in comet
composition can tell us about the relative formation conditions of
these comets.

Protostellar disks provide complementary information to the comet
observations. Their chemistry is complex, covering ice chemistry in the cold,
shielded midplane, warm gas phase chemistry and hot, highly ionized
chemistry.  Comets form from the ices in the midplane, yet 
dynamical processes ensure that this region does not evolve in
isolation but rather is the end product of material that has been
processed in a variety of different environments and then brought
together in the cometary ices.   Establishing the origin of cometary
ices therefore requires not only studying the midplane but also
an understanding the chemical and physical evolution of the disk
as a whole.  By studying disks we can observe the chemistry and
physics that shaped the solar system as they shape other
protoplanetary systems. 

Molecular deuteration is a link that might tie
together the disparate objects of the solar system and their formation
history and give us the picture of the whole solar system formation
history. However, at present this link has still weak connections that
need to be reinforced. The advent of the powerful new facilities, as
for example ALMA and IRAM-NOEMA, will certainly allow this field to
progress rapidly in the next few years.

Rosetta will provide detailed ground-truths for the
 organic and isotopic makeup of cometary volatiles.  The new insights
 obtained into the chemical processes which likely produced these
 molecules -- interstellar or nebular -- will allow us to constrain
 models of the formation and evolution of the early solar system.
Its instruments will characterize the composition of the
comet 67P/Churyumov-Gerasimenko and distinguish between the molecules native
to the comet nucleus (parent species) and those generated by
chemistry or by fragmentation in the comae (daughter species).  The
observations it will make will provide new insights into the formation
conditions of the comet.  They will also elucidate the degree to which interstellar material
can survive its journey into
the solar nebula and from there into comets.

Several measurements will characterize the its thermal history, e.g.,
measurements of molecular spin temperatures \citep{rosetta} and deuteration and of the
abundance of argon \citep{hassig}. 
The relative abundances of elements
can tell us something about the dominant molecular form of each
element at the time of formation -- if an element is depleted relative
to others then the major carrier of that element must have been too
volatile to be condensed at the time of formation, or must have
desorbed in the intervening time.  In particular Rosetta measurements
may help to elucidate the mystery of the lack of nitrogen in comets
 \citep{altwegg14}.

A key measurement will be the ratio of oxygen isotopes in water.  Meteoritic
water has a very different ratio of the oxygen isotopes compared to
oxygen isotopes in the Sun.  Rosetta provides our best opportunity to
directly detect the isotope ratio from the early solar system, and in
so doing will distinguish between the various theoretical models of
how the oxygen isotope fractionation occurred.

In summary, Rosetta's observations will provide new
insights into the chemical processes that resulted in the formation of
cometary volatiles and consequently into the conditions that were
present in the solar nebula at their time of formation.  

\begin{acknowledgements}
The work of K. Willacy and C. Alexander was carried out at the Jet Propulsion
Laboratory, California Institute of Technology, under a contract with
the National Aeronautics and Space Administration.  S. B. Charnley and
S. N. Milam acknowledge the support of the Goddard Center of Astrobiology.  Support
for K. Willacy, S. B. Charnley and S. N. Milam was partially provided by the NASA Origins of Solar
Systems Program.  E. Gibb was supported by the NASA Exobiology and
Evolutionary Biology (grant NNX11AG44G) and NSF Planetary Astronomy
(grant AST-1211362) programs. 
 O. Mousis and C. Ceccarelli acknowledge support
from CNES.  O. Mousis thanks the support of the A*MIDEX project
(no. ANR-11-1DEX-0001-02) funded by the ``Investissements d'Avenir''
French government program managed by the French National Research
Agency (ANR).  M. Ali-Dib was supported by a grand from the city of Bean\c{c}on.
E. S. Wirstr\"om was supported by the Swedish National Space Board. 

\noindent\textcopyright 2014 Jet Propulsion Laboratory, California Institute
of Technology.  All rights reserved.

\end{acknowledgements}

\bibliographystyle{spr-chicago}      


\end{document}